\documentclass[footinbib,aps,pra,reprint,superscriptaddress]{revtex4-1}

\usepackage{amsmath,amssymb,braket,upgreek,bm,verbatim}
\usepackage{graphicx,tabularx} 
\usepackage[colorlinks=true,bookmarks=false,citecolor=blue,urlcolor=blue]{hyperref} %
\usepackage{xcolor}

\usepackage[absolute]{textpos}
\usepackage{xcolor}
\usepackage{multirow}
\newcommand\fix[1]{\textcolor{red}{#1}}

\def\bx{\mathbf{x}}
\def\by{\mathbf{y}}

\def\bbD{{\bm{\mathcal{D}}}}

\def\cF{\mathcal{F}}

\DeclareMathOperator{\Tr}{Tr}

\usepackage{amsmath,amssymb,bbm,bm}
\usepackage{mathtools}
\usepackage{braket}
\usepackage{amsmath}
\usepackage{relsize}
\usepackage{comment}
\usepackage{csquotes}

\def\bx{\mathbf{x}}

\def\bbD{{\bm{\mathcal{D}}}}

\newcommand{\appropto}{\mathrel{\vcenter{
  \offinterlineskip\halign{\hfil$##$\cr
    \propto\cr\noalign{\kern2pt}\sim\cr\noalign{\kern-2pt}}}}}

\begin{document}
\title{Nonlocal subpicosecond delay metrology using spectral quantum interference}

\author{Suparna Seshadri}
\author{Navin Lingaraju}
\author{Hsuan-Hao Lu}
\affiliation{School of Electrical and Computer Engineering and Purdue Quantum Science and Engineering Institute, Purdue University, West Lafayette, Indiana 47907, USA}
\author{Poolad Imany}
\affiliation{National Institute of Standards and Technology, Boulder, CO 80305, USA}
\affiliation{
Department of Physics, University of Colorado, Boulder, CO 80309,USA}

\author{Daniel E. Leaird}
\author{Andrew M. Weiner}
\affiliation{School of Electrical and Computer Engineering and Purdue Quantum Science and Engineering Institute, Purdue University, West Lafayette, Indiana 47907, USA}

\date{\today}

\maketitle


\textbf{Timing and positioning measurements are key requisites for essential quantum network operations such as Bell state measurement. Conventional time-of-flight measurements using single photon detectors are often limited by detection timing jitter. In this work, we demonstrate a nonlocal scheme to measure changes in relative link latencies with subpicosecond resolution by using tight timing correlation of broadband time-energy entangled photons. Our sensing scheme relies on spectral interference achieved via phase modulation, followed by filtering and biphoton coincidence measurements, and is resilient to microsecond-scale mismatch between the optical link traversed by the biphotons. Our experiments demonstrate precision of $\pm$$0.04$ ps in measurements of nonlocal delay changes and $\pm$$0.7^{\circ}$ in measurements of radio-frequency phase changes.  Furthermore, we complement our technique with time-tag information from single photon detectors in the same setup to present unambiguous sensing of delay changes. The proposed technique can be implemented using off-the-shelf telecom equipment thus rendering it adaptable to practical quantum network infrastructure.}


\section{Introduction}
Broadband time-energy entangled photons, by virtue of their strong correlations in temporal and spectral degrees of freedom, have exhibited high utility for developing quantum technologies like quantum communication \cite{wengerowsky2018entanglement, ursin2007entanglement, honjo2008long, yin2020entanglement, nunn2013large}, sensing \cite{SQLlloyd,chen2019hong, DistPhase}, spectroscopy \cite{lee2019interferometric, yabushita2004spectroscopy}, positioning and clock synchronization \cite{clklee2019symmetrical, giovannetti2001clock,ho2009clock, giovannetti2001quantum}. Multi-node quantum networks with the ability to distribute entanglement over multiple length scales–-potentially across the world, will be key enablers in advancement towards distributed quantum systems \cite{DistPhase,NetOfClocks} for enhanced computing, sensing and long distance secure communication. In this endeavor, accurate timing and positioning metrological measurements will play a crucial role in realizing essential network operations such as switching and routing. As local and metropolitan area fiber-optic quantum networks evolve from entanglement distribution \cite{spiropulu2021illinois, alshowkan2021reconfigurable, lingaraju2021adaptive} to heralded entanglement generation \cite{barz2010heralded, scherer2011long, sun2017entanglement}, ability to sense variations in link latency will be prerequisite to critical network tasks such as Bell state measurement.


There have been previous investigations on measuring temporal correlations of entangled photons via coincidence measurement, purposed for high-precision synchronization between remote sites \cite{ho2009clock,nonlocalquan2020high,giovannetti2001clock}.
Entanglement offers a distinguishing non-local resource \cite{fransonnonlocal,fitchfransondispersion, chromdisp, nonlocalmod} that can be exploited in the quantum network architecture. The nonlocal measurement of temporal correlation is often addressed via tagging the biphoton arrival times at remote sites using single photon detectors (SPDs) and event timers. A simple time-resolved correlation measurement using commercially available superconducting nanowire SPDs is however limited by timing jitters on the order of 50 ps. There have been demonstrations to push the limits of temporal sensitivity both by improving the SPD hardware \cite{Sub3pskorzh2020demonstration} and by modifying the event timing algorithms \cite{nonlocalquan2020high}. However, nonlocal biphoton delay measurement with a sensitivity independent of detector/event timer resolution has so far not been reported. While interferometric sensing techniques such as Hong-Ou-Mandel \cite{lyons2018attosecond,chen2019hong} and Ramsey interferometry \cite{clemmen2016ramsey} are not limited by detection timing jitter, they do not offer the nonlocal sensing capability. Hong-Ou-Mandel interferometry relies on spatial overlap of the photons requiring balancing of path lengths traversed by them. In the case of Ramsey interferometry, nonlinear interaction between the photons becomes crucial for realization.

\begin{figure*}[!htb]

  \centering
  \includegraphics[width=\textwidth]{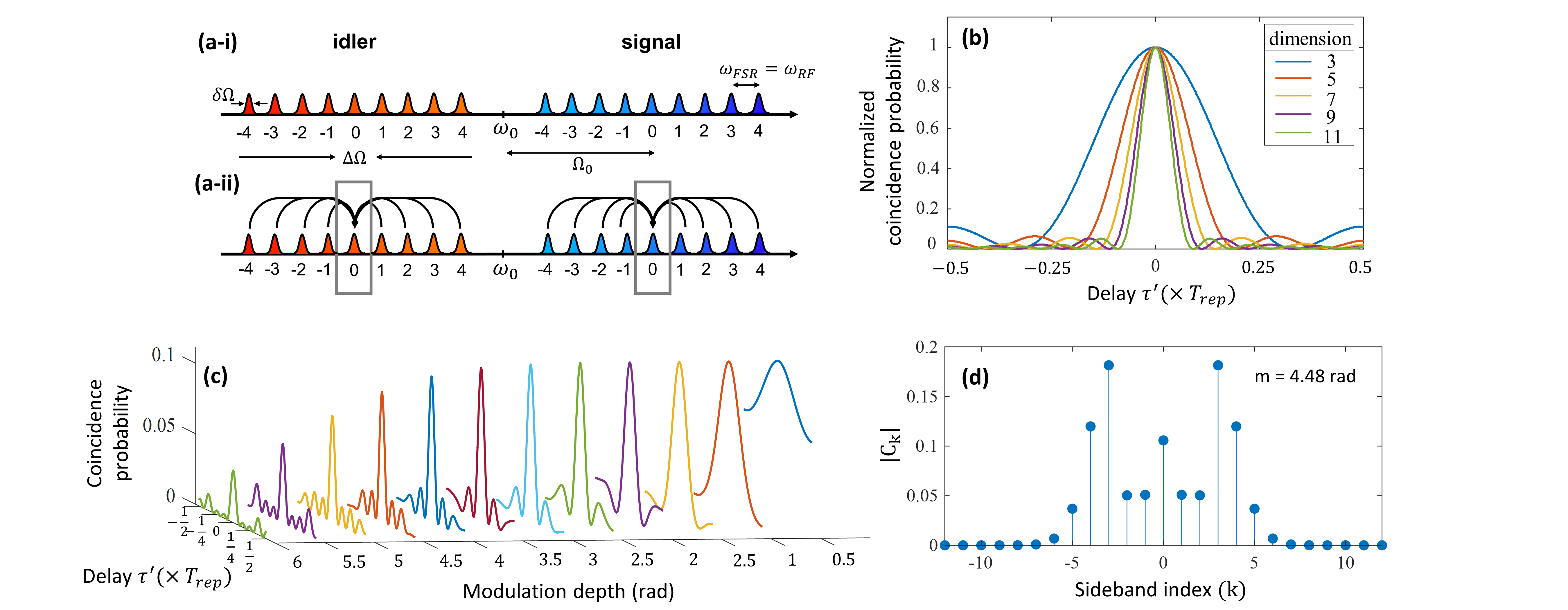}
\caption{
(a-i) Frequency domain schematic of a nine dimensional BFC.
(a-ii) Illustration of phase modulation sidebands contributing to the bin pair $\ket{0,0}_{SI}$, selected for coincidence detection. (b) Theoretical coincidence probability from equal-amplitude mixing of frequency bins plotted over one repetition period ($T_{rep}$) for BFCs with different dimensions.  
(c) Theoretical coincidence probability per photon pair for nine-dimensional BFC resulting from single sine-wave phase modulation at different modulation depths. (d) Bessel mixing coefficients at modulation depth of 4.48 radians.}

\label{Theory_fig}
\end{figure*}

Here, we report the first proof-of-concept demonstration of nonlocal time-delay sensing independent of SPD jitter and resolution of timing electronics. 
This scheme also offers potential for measuring relative radio frequency (RF) phase drifts in sinusoidal phase modulation applied along different fiber links. Our approach is inspired from previous studies \cite{olislager2010frequency,kues2017chip,imany201850, imany2018characterization, imany2020probing} that include the characterization of spectral phase coherence in high dimensional frequency bin entangled biphotons using phase modulation and spectral filtering operations. 
Building on these works, here we propose and demonstrate a derived scheme, for nonlocal sensing of temporal delay and RF phase drifts without the need for low jitter SPDs. 
Further, we show the capability for measuring subpicosecond variations between delays of two fiber links without requiring to balance their path lengths. In this scheme [cf.~Fig.~\ref{setup}], time-energy entangled signal and idler photons 
are routed through different fiber links with nonidentical delays. Down the links, the photons undergo sinusoidal electro-optic phase modulation followed by spectral filtering, employed to select correlated frequency bins for coincidence detection. The relative phase drifts between the RF sinusoids modulating the biphotons as well as the relative path length changes in the links effectively contribute to the 
joint spectral phase of the biphoton frequency bin pairs. Owing to the coherent frequency mixing from the sinusoidal phase modulation, the spectral phase on the biphotons is mapped to the coincidences measured between the 
selected frequency bins.  
The interferogram measured from this approach relies only on the total coincidences acquired by the SPDs; thus, the detection timing jitter does not limit the delay sensitivity. 

Various characteristics and prospects of the proposed scheme are reported in this work.  We first demonstrate the nonlocal sensing potential of this approach using the coincidence-interferogram, acquired 
as a function of relative delay changes and RF modulation phase drifts between different fiber links.  Second, we employ dispersion compensation to demonstrate uncompromised delay sensitivity despite several meters of length mismatch between the two optical links. Unlike most prevalent interferometric sensing techniques, the proposed approach does not require precise balancing of optical paths travelled by the biphotons. 
Although the biphoton-coincidences in this scheme are periodic in delay due to sinusoidal phase modulation, we successfully resolve the associated delay ambiguity using photon arrival time-tags from SPDs as a coarse measurement. Lastly we present theoretical and experimental evidence showing strong parallels between the second-order time-correlation of the biphotons and the coincidence measurements acquired from the proposed sensing technique.

The utility of our sensing scheme in quantum networks can conceivably range from tracking link latency and relative RF phase drifts to distant clock synchronization and position verification \cite{clklee2019symmetrical,clkvalencia2004distant,giovannetti2001clock,giovannetti2001quantum}. Although we utilize programmable filters for spectral demultiplexing and filtering operations, a deployable system can be realized by replacing them with coarse and dense wavelength division multiplexers. The proposed sensing approach is potentially adaptable in practical quantum networking applications since all the equipment needed for these measurements are expected to be staples of fiber-based quantum networks. 


The rest of the article is organized as follows. Section \ref{theory} presents the theoretical analysis of the proposed spectral interferometric approach and its dependence on biphoton delay and RF modulation phase. In section \ref{experiment}, we discuss the experimental apparatus and associated results revealing key features of this scheme. Finally, in section \ref{discussion} we summarize and discuss ideas on adapting our work for practical quantum networking applications.

Preliminary results related to sections \ref{ cleo1 } and \ref{1.5us_mismatch} have been reported in conferences \cite{seshadri2020precision} and \cite{seshadri2021nonlocal}. The current manuscript presents additional results, including new experiments demonstrating unambiguous delay sensing using complementary time-tag information and elucidating the relationship with the biphoton time-correlation function, as well as a comprehensive theoretical analysis.

\section{Theory}\label{theory}

Consider the state of a biphoton frequency comb (BFC) of dimension 2N+1, with a free spectral range (FSR) of $\omega_{_{FSR}}$ represented as 
\begin{equation}\label{SS1}
\ket{\Psi} =\sum_{k=-N}^{ N} \alpha_k \big[\hat{a}^{(S)}_{k}\big]^\dagger \big[\hat{a}^{(I)}_{-k}\big]^\dagger \ket{\mathrm{vac}},
\end{equation} where $\ket{\mathrm{vac}}$ is the vacuum state, $\alpha_k$ is the complex probability amplitude of the frequency bin pair associated with the creation operators $\big[\hat{a}^{(S)}_{k}\big]^\dagger$ and  $\big[\hat{a}^{(I)}_{-k}\big]^\dagger$ corresponding to the $k^{th}$ signal bin and $-k^{th}$ idler bin centered at $\omega^{(S)}_{k} = \omega_0 + \Omega_0 + k\omega_{_{FSR}} $ and $\omega^{(I)}_{-k} = \omega_0 - \Omega_0 - k\omega_{_{FSR}}$ respectively, with $\Omega_0$ being the frequency offset of the signal and idler spectra from the center frequency $\omega_0$. An illustration of the BFC described above with dimensionality of nine is shown in Fig.~\ref{Theory_fig}(a). 

The signal and idler frequency bins are routed to different optical links using a pulse shaper that is also employed to impart a spectral phase on the biphotons. The signal and idler traverse through delays $\tau_{_S}$ and $\tau_{_I}$ respectively along their paths. 
Down the links, the biphotons are each phase modulated with RF sinusoids at modulation frequency $\omega_{_{RF}}$ equal to the bin spacing $\omega_{_{FSR}}$. Finally, frequency bins centered at $\omega^{(S)}_{0} = \omega_0 + \Omega_0$ and $\omega^{(I)}_{0}  = \omega_0 - \Omega_0$, (i.e., when $k = 0$) are selected using spectral filters and routed to different SPDs for coincidence detection.

The annihilation operator $\hat{a}^{(S)}_{k}$ ($\hat{a}^{(I)}_{-k}$) corresponding to the signal (idler) transforms into $\hat{b}^{(S)}_{k}$($\hat{b}^{(I)}_{-k}$) after traversing down the link (prior to phase modulation) as follows:

\begin{equation}\label{operatorS}
  \hat{b}^{(S)}_{k} = \Big[\hat{a}^{(S)}_{k}\Big] \Big[\exp{\big(i\tau_{_S}\omega^{(S)}_{{k}}\big)}\Big] H^{(S)}_k,
\end{equation}

\begin{equation}\label{operatorI}
  \hat{b}^{(I)}_{-k} = \Big[\hat{a}^{(I)}_{-k}\Big] \Big[\exp{\big(i\tau_{_I}\omega^{(I)}_{{-k}}\big)}\Big] H^{(I)}_{-k},
\end{equation} where $H^{(S)}_{k}$ and $H^{(I)}_{-k}$ represent the frequency dependent complex amplitudes (i.e frequency dependent amplitude and phase) that $k^{th}$ signal and $-k^{th}$ idler bins pick up prior to the phase modulation (in addition to temporal delay). Such effects can be applied intentionally using the pulse shaper or arise from dispersive fiber propagation, etc. For more details on our sign convention for delay and other phases, see \textit{Supplement A}. We first consider linear spectral phase ramps applied by the pulse shaper on the signal and idler bins, given by $H^{(S)}_{k} =e^{ ik \varphi_{_S}}$ and $H^{(I)}_{-k} = e^{-ik \varphi_{_I}}$.
 
 Down the links, after the transformation due to temporal phase modulation of the form $e^{-im \sin(\omega_{_{FSR}} t + \phi_{_{S}})}$ and $e^{-im \sin(\omega_{_{FSR}} t + \phi_{_{I}})}$ respectively applied in the signal and idler paths, the annihilation operators $\hat{c}^{(S)}_{k}$ and $\hat{c}^{(I)}_{-k}$ corresponding to signal at $k^{th}$ frequency bin and idler at $-k^{th}$ frequency bin are given by

\begin{equation}
\begin{aligned}
\label{EfieldS}
\hat{c}^{(S)}_{k} = \sum_{p=-\infty}^\infty J_{p}(m) e^{-ip\phi_{_S}} \Big[\hat{b}^{(S)}_{k-p}\Big], 
\end{aligned}
\end{equation}

\begin{equation}
\begin{aligned}
\label{EfieldI}
\hat{c}^{(I)}_{-k} = \sum_{q=-\infty}^\infty J_{q}(m) e^{-iq\phi_{_I}} \Big[\hat{b}^{(I)}_{-k-q}\Big], 
\end{aligned}
\end{equation} where $J_l(m)$ is the Bessel function of the first kind of integer order $l$, $m$ is the modulation depth in radians,  $\phi_{_S}$ and $\phi_{_I}$ are the phases of the RF sinusoidal waveforms modulating the signal and idler photons respectively.

Based on the above formalism, the probability of measuring a coincidence count, $ \mathcal{P}(\Delta\phi)$ between the frequency bins at $\omega^{(S)}_{0}$ and $\omega^{(I)}_{0}$, per input photon pair, is given by the following equation:

\begin{equation}\label{SS5}
\begin{aligned}
\mathcal{P}(\Delta\phi) &= \Big|\bra{\mathrm{vac}} \hat{c}^{(S)}_{0} \hat{c}^{(I)}_{0} \ket{\Psi}  \Big|^2\\
&\propto \left| \sum_{k=-N}^{N}  \alpha_k C_k e^{i k\Delta\phi}  \right|^2
\end{aligned}
\end{equation} where,
\begin{equation}\label{SS6}
\begin{aligned}
&\Delta\phi = \omega_{_{FSR}}\tau + \phi_{_{RF}} + \varphi_{_{PS}}
&\tau = \tau_{_S} - \tau_{_I}&\\
&\phi_{_{RF}} = \phi_{_S} - \phi_{_I} 
&\varphi_{_{PS}} = \varphi_{_S} - \varphi_{_I}\\
\end{aligned}
\end{equation}
Here, the mixing coefficient $C_k = J_k(m)J_{-k}(m) = {|J_k(m)|^2}e^{ik\pi}$ results from the phase modulation sidebands.  For more details on the theory incorporating frequency bin width and detection timing jitter, see \textit{{Supplement A}}. We observe that the coincidence probability is sensitive to the differential biphoton delay $\tau$, the relative phase $\phi_{_{RF}}$ between the RF drive signals to the phase modulators, and the linear spectral phase increment $\varphi_{_{PS}}$ imparted by pulse shaper. By collecting all the linear spectral phase terms, we can rewrite the coincidence probability per photon pair as 

\begin{equation}\label{SS7}
   \begin{aligned}
&\mathcal{P}(\tau') \propto \left| \sum_{k=-N}^{N}  \alpha_k |C_k| e^{ik\omega_{_{FSR}}\tau'}  \right|^2,\\
\end{aligned}
\end{equation} where $\tau' = (\Delta\phi + \pi)\omega_{_{FSR}}^{-1}$ and $|C_k| = |J_k(m)|^2$. It is evident from the above analysis that 
the coincidence probability is periodic with respect to the effective differential biphoton delay $\tau'$, with a repetition period  given by the inverse of free spectral range i.e, $T_{rep } = {2\pi}\omega_{_{FSR}}^{-1}$. Similarly the coincidence probability repeats every $2\pi$ radians with respect to incremental biphoton phase $\Delta\phi$.


It is worth noting that the scheme outlined above is similar to that which has been used to characterize frequency bin entanglement \cite{olislager2010frequency,kues2017chip,imany201850,imany2018characterization}. The difference is that here for the first time, we focus on the dependence of both the differential biphoton delay and RF phase.  Our results show that frequency bin quantum interference can be exploited for sensing of these quantities.

Figure~\ref{Theory_fig}(b) plots the coincidence probabilities over one period under equally weighted mixing ( i.e., $|C_k| = 1$) and uniform probability amplitudes (i.e, $\alpha_k = 1/\sqrt{{2N+1}}$) as a function of $\tau'/T_{rep}$. With higher dimensionality, the width of the trace in Fig.~\ref{Theory_fig}(b) decreases and the maximum slope in the main lobe of the trace increases, offering better sensitivity with respect to changes in delay and RF phase. Under this ideal equally weighted mixing scenario, the width of the coincidence trace is inversely proportional to the dimensionality ($2N+1$). However, in actual experiments the mixing coefficients are not equal; they have a Bessel function dependence on the phase modulation amplitude. To provide further insight, in Fig.~\ref{Theory_fig}(c) we plot theoretical coincidence traces for Bessel mixing coefficients as described in Eqs.~(\ref{EfieldS}-\ref{SS7}) over modulation depths ranging from 0.5 to 6 radians (experimentally feasible using a single phase modulator). Here we assume a nine dimensional BFC, which coincides with our experiments, and retain equal probability amplitudes $\alpha_k$. 
We observe that the widths of the traces decrease with increasing modulation depth up to about 4 rad, while the coincidence probability at the peak remains roughly constant.  For higher modulation depths, the peak coincidence probability decreases while the widths of the traces remain approximately constant in the considered range going upto 6 radians.  We can understand these trends as follows: As the modulation depth increases, the phase modulators generate sidebands over a wider bandwidth, with decreased amplitude per sideband. As a result the number of frequency bins contributing to the central ($k=0$) bins selected for two photon interference increases, leading to a higher effective dimensionality and narrower traces. However, for sufficiently high modulation depth ($m \gtrsim 4$), the number of sidebands exceeds the number of frequency bins in the initial BFC (nine in our example). Now the effective dimensionality is limited by the number of frequency bins and the widths of the traces depend only on the distribution of the $\pm 4$ sideband amplitudes. The increased modulation depth now results in an effective loss, since sidebands are generated outside the frequency space in which they can be used.

In our experiments, the depth of phase modulation on both the photons is set to be $\sim$ 4.48 rad (with magnitude of sideband intensities shown in Fig.~\ref{Theory_fig}(d)), serving to operate near optimal delay sensitivity setting for a nine dimensional BFC with a constant input flux rate; see \textit{Supplement B} for more details.


\section{Experiment and Results}\label{experiment}

\subsection{Setup}
\begin{figure}[!htb]
\centering
\fbox{\includegraphics[width=\linewidth]{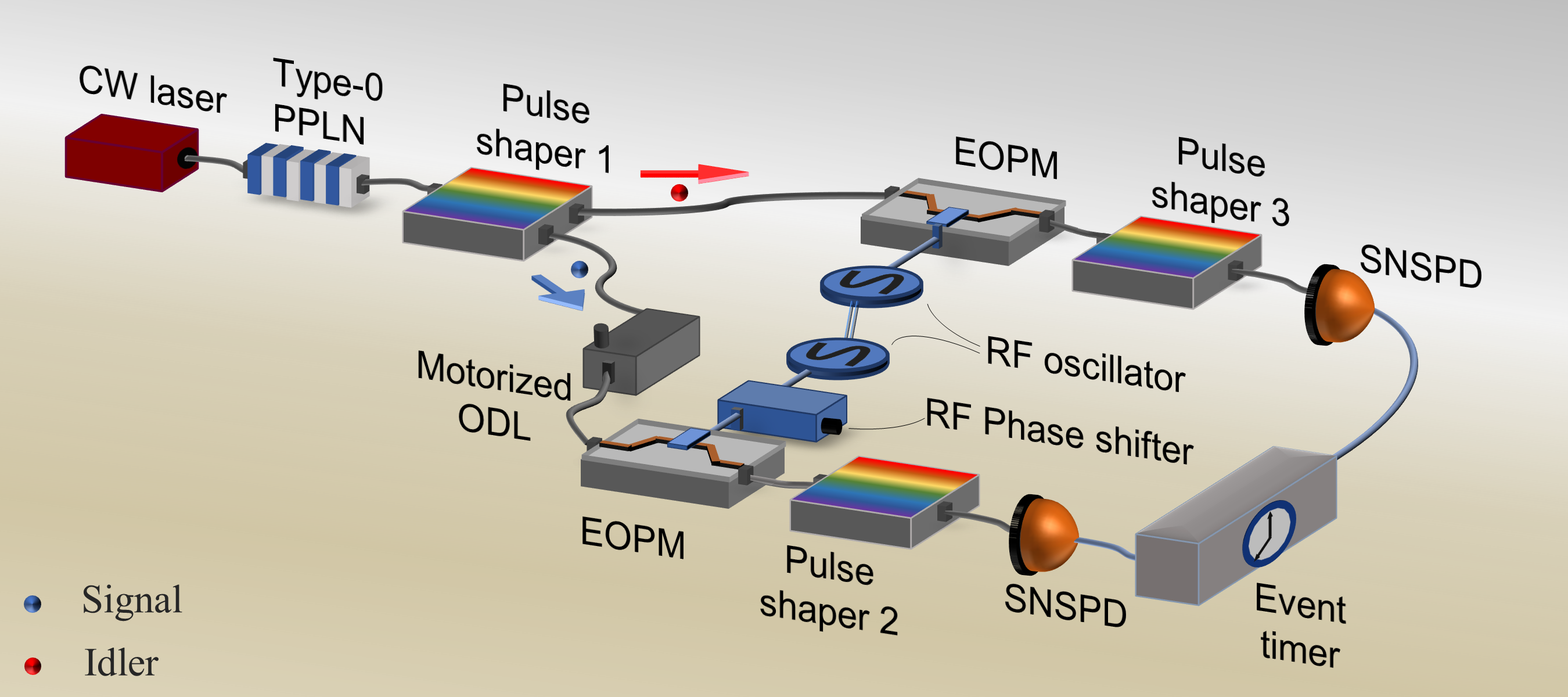}}
\caption{The experimental setup. CW laser: continuous-wave laser, PPLN: periodically poled lithium niobate waveguide, ODL: Optical delay line, EOPM: Electro-optic phase modulator, SNSPD: superconducting nanowire single photon detector. 
}
\label{setup}
\end{figure}

In this section we present some proof-of-concept experiments, demonstrating the key capabilities of our sensing approach. As sketched in Fig.~\ref{setup}, a continuous-wave pump laser at 778 nm is routed to a 2.1-cm-long fiber-pigtailed periodically poled lithium niobite (PPLN) ridge waveguide to generate time-energy entangled photons under type-0 phase matching. The spontaneous parametric downconversion (SPDC) spectrum is centered around 1556 nm  ($\omega_0= 2\pi \times192.7$ THz) and spans a bandwidth $>$ 5 THz. The 778 nm pump laser 
used in the setup has a linewidth of $\sim$200 kHz (corresponding to biphoton coherence length of $\sim$1 km). We use a programmable pulse shaper (Pulse shaper 1) to  select frequency-correlated slices of spectral width 
$\sim$288 GHz from the signal and idler spectra and route them to different arms. The centers of selected signal and idler spectral slices are offset from the SPDC center frequency by {$\Omega_0/2\pi = $ 608 GHz}. The biphotons travel through different path lengths as they propagate along their respective optical links. Down the link, an electro-optic phase modulator (EOPM) is placed in each of signal and idler arms. The EOPMs are driven by RF sinusoidal waveforms at a modulation frequency \mbox{$\omega_{_{RF}}/2\pi = $ 32 GHz} and modulation depth $m \approx $ 4.48 radians. Our scheme entails the synchronization of RF waveforms driving the modulators in order to ensure phase coherent interaction of the frequency bin pairs in the biphoton spectrum. In our experiment, the two RF waveforms modulating the biphotons are derived from a common signal generator. 

\begin{figure*}[!htb]
  \centering
  \includegraphics[width=\textwidth]{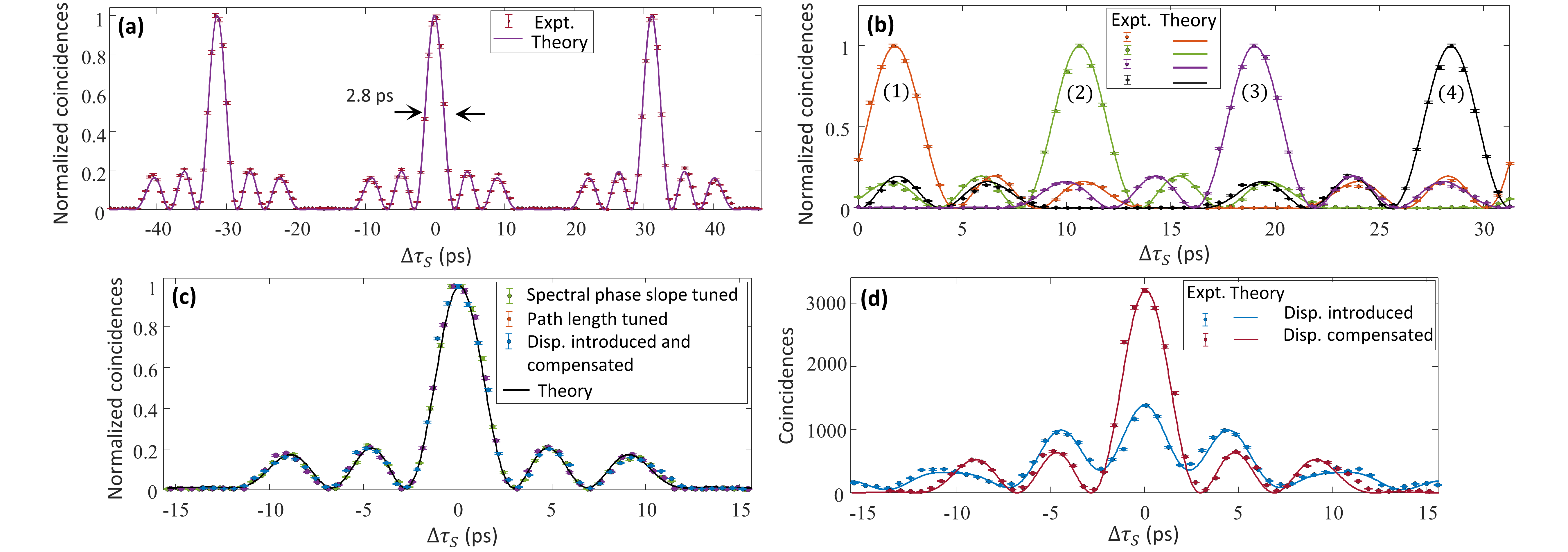}
\caption{Two-photon coincidence interferograms acquired as a function of change in group delay of the signal photon $\tau_{_S}$. Signal photon delay scanned by the ODL (a) over three repetition periods, (b) after each turn of the RF phase shifter control knob (phase shift $\sim100^\circ$ per turn). (c) Overlaying three normalized coincidence-interferograms that are acquired as follows: (i)  by varying the slope of linear spectral phase on the signal bins using Pulse shaper 1, (ii - iii) by scanning the signal photon delay using an ODL in the absence of additional dispersive fiber \& after dispersion compensation in the  presence of a dispersive spool in the signal arm respectively. (d) Coincidences (acquired over 10 s) plotted as the signal photon delay is scanned in the presence of dispersive spool in the signal arm, before and after dispersion compensation.}

\label{3Rep_RF}
\end{figure*}

After phase modulation, frequency bins with spectral width of $\delta\Omega = 15$ GHz are selected from the centers of the signal and idler spectra using two programmable filters (Pulse shapers 2 and 3, respectively). The selected frequency bins are routed to superconducting nanowire single-photon detectors (SNSPDs) for coincidence measurement. In our demonstration we use SNSPDs with combined timing jitter of $\sim$ 100 ps and an event timer to record coincidence time-tag histograms with a resolution upto $\sim$1 ps. 

The phase modulation and spectral filtering operations effectively post-select a 
a nine dimensional BFC state, with FSR equal to the modulation frequency i.e $\omega_{_{FSR}} = \omega_{_{RF}}$, and frequency bin width \mbox{equal to $\delta\Omega$}; see detailed theoretical analysis in \textit{Supplement A}. The selected frequency bin pair contains coherent sideband contributions from all the nine frequency bins accommodated in the biphoton spectra. Thus the probability of coincidence detection depends on both the spectral phase across the nine bin pairs and the modulation parameters. 

Note that although the parameter settings specified above are used in the initial experiments,  e.g., Figs.~\ref{3Rep_RF}(a-b), we alter them in some of the later demonstrations. The total acquisition time of the coincidence histograms ($\Delta t$), the histogram-time window ($\Delta T$) over which coincidences are integrated to plot the datapoints in the figures, bandwidth of the signal and idler spectrum ($\Delta \Omega$), offset from the SPDC center ($\Omega_0$), modulation frequency ($\omega_{_{FSR}}$) and the frequency bin width ($\delta\Omega$) used in all the experiments are tabulated in \textit{Supplement C}. The dimensionality of the BFC (d=9) and the phase modulation depth (m $\sim$ 4.48 rad) are set to be the same across all the experiments.

\subsection{Interferograms from detuning differential biphoton delay}\label{ cleo1 }

Figure~\ref{3Rep_RF}(a) shows coincidence counts measured when the delay of the signal photon $\tau_s$ is varied using the optical delay line.  The data show a series of sharp peaks that repeat at $\sim$31.25 GHz, corresponding to the period of the 32 GHz RF modulation. The full width at half maximum (FWHM) of the interferogram is equal to 2.8 ps, far below the $\sim$100 ps timing jitter of our SNSPDs. We can clearly resolve subpicosecond delay steps especially at operating points situated close to the high-slope regions in the central lobe of the interferogram.


The relative delay between the biphotons is detuned using a motorized optical delay line (ODL) placed in the signal arm prior to the EOPM, as shown in Fig.~\ref{setup}. 
The ODL is scanned in increments of 0.55 ps. The acquired-coincidence histogram is normalized with respect to the peak and fit with the theoretical probability in Eq.~(\ref{SS7}) considering bin pairs with uniform probability amplitudes \mbox{(i.e., $\alpha_k = 1/\sqrt{2N+1}$)}. The change to the group delay of the signal photon is denoted by $\Delta\tau_{_S}$ in the plots.

 Figure~\ref{3Rep_RF}(a) shows close agreement between the theoretical prediction and the experimental results. The dispersion accumulated by the biphotons ($\sim$ 15 m SMF in each arm) is ignored since it is negligible in broadening the interferogram in comparison to the statistical error in the delay. 


 In our next result, Fig.~\ref{3Rep_RF}(b) plots a series of interferograms acquired at four different RF phase settings of the 32-GHz waveform modulating the signal photon. The shifts in the interferograms corroborate the sensitivity to relative RF phase as predicted by Eq.~(\ref{SS7}). The RF phase shifter is manually adjusted in steps of roughly $\sim 100^\circ$ and at each RF phase setting, the motorized ODL is scanned.
 Here, we estimate the relative RF shifts from a least square fit of the interferograms to the theory. From Eq.~(\ref{SS5}), if the RF modulation phase shift of trace (1) with respect to trace (2) is $\phi_{_{RF}}^{(1)} - \phi_{_{RF}}^{(2)}$, then the resultant offset in ODL setting corresponding to the peaks of the traces (1) and (2) is 

\begin{equation}
\tau^{(1)} - \tau^{(2)} = - \big(\phi_{_{RF}}^{(1)} - \phi_{_{RF}}^{(2)}\big)\omega_{_{FSR}}^{-1}. 
\end{equation}\label{RF_delay}

The RF phase settings corresponding to the interferogram (1) relative to interferograms (2-4) are recovered to be :  $\phi_{_{RF}}^{(1)} - \phi_{_{RF}}^{(2)} =$ $102.0^{\circ} \pm  0.7^{\circ}$, $\phi_{_{RF}}^{(1)} - \phi_{_{RF}}^{(3)} =$ $198.6^{\circ} \pm 0.6^{\circ}$, and $\phi_{_{RF}}^{(1)} - \phi_{_{RF}}^{(4)} =$ $307.0^{\circ} \pm 0.6^{\circ}$; the reported 95\% confidence bounds incorporate the errors from the least squares fit of the interferograms and the resolution of the ODL.


Equation~(\ref{SS6}) predicts that the two photon interferograms are equally sensitive to delays from changes in the physical path length and to delays generated without moving parts by application of a linear spectral phase. Figure~\ref{3Rep_RF}(c) overlays one period of the interferogram from Fig.~\ref{3Rep_RF}(a) over that measured by scanning the slope of linear spectral phase on the signal frequency bins using Pulse shaper 1 (while no phase is applied on idler bins). From Eq.~(\ref{SS7}), a spectral phase increment of $\varphi_{_S}$ applied to the signal frequency bins is  associated with relative delay $\tau'$ modulo $T_{rep}$ given by {$\varphi_{_S}\omega_{_{FSR}}^{-1}$}. This equivalence of relative delay with linear spectral phase is evident from the closely matching traces in Fig.~\ref{3Rep_RF}(c). 
The interferogram in our proposed scheme can thus be measured without requiring an optical delay line, by simply using a pulse shaper already employed in our setup for demultiplexing the signal and idler photons. 

Together, Figs.~\ref{3Rep_RF}(a-c) confirm the sensitivity of our two photon interferograms to changes in the relative physical delays of signal and idler, to changes in the relative phases of the applied RF modulations, and to changes in applied linear spectral phases, as predicted by the $\Delta\phi$ expression in Eq.~(\ref{SS6}). {Fig.~\ref{3Rep_RF}(c) also plots an interferogram acquired after compensating for the total dispersion in the setup when an SMF spool introducing $\sim$ 1.5 $\mu$s of delay is inserted in the signal arm. The details are covered in the following section.}


\subsection{Sub-ps sensitivity over $\mu s$-scale delay mismatch}\label{1.5us_mismatch}

The proposed scheme does not necessitate the precise balancing of path lengths traversed by the biphotons in order to observe the interferogram. The interference features can be observed as long as the imbalance in the signal and idler arms is within the biphoton coherence length (dictated by inverse of the CW pump linewidth). A characteristic feature of the interference trace is its periodicity due to the sinusoidal phase modulation. One can measure small delay changes (modulo the modulation period) even with large delay offsets, because the delay changes are measured with respect to a periodic RF clock.  
In this section, we demonstrate sub-ps resolving capability despite large imbalances in the path lengths in the signal and idler arms. A $\sim$313-meter-long SMF-28e spool is inserted in the signal path prior to modulation, which introduces $\sim$ 1.5 $\mu$s delay offset between the biphotons arriving at their respective detectors.
Figure~\ref{3Rep_RF}(d) shows the coincidences measured (in blue markers) 
at each ODL setting as delay in the signal path is scanned. 
The trace still achieves picosecond scale delay sensitivity but is broadened and distorted due to dispersion. The total width of the 
interferogram at the half maximum points is 10.8 ps, close to four times that in the absence of the SMF spool. 

By factoring in second-order dispersion into the theoretical coincidence probability, Eq.~(\ref{SS7}) becomes

\begin{equation}\label{EqDispNL}
\mathcal{P}(\tau_{_{eff}}) \propto \Bigg| \sum_{k=-N}^{ N} |C_k|   e^{\big( \frac{1}{2}i\beta_2 (L_S+L_I)k^2 \omega^2_{_{FSR}} +ik\omega_{_{FSR}}\tau_{_{eff}} \big) } \Bigg|^2
\end{equation}

\begin{equation}\label{EqDispNL_2}
\begin{aligned}
\tau_{_{eff}} = \tau &+  \beta_2(L_S+L_I)\Omega_0+\omega_{_{FSR}}^{-1}(\phi_{_{RF}}{+\varphi_{_{PS}}} + \pi ) 
\end{aligned}
\end{equation} where $\beta_2 $ is the dispersion parameter of the fiber, $\tau_{_{eff}}$ is the effective differential biphoton delay in the presence of dispersion, $L_S$ and $L_I$ are the respective fiber lengths over which signal and idler accumulate dispersion. See \textit{{Supplement A.1}} for details on the derivation.

Although dispersive broadening is undesirable, its effect can be accurately modelled, and it can be compensated. We fit Eq.~(\ref{EqDispNL}) to the interference pattern in Fig.~\ref{3Rep_RF}(d)(blue) and estimate the sum total of dispersion $\beta_2(L_S+L_I)$ associated with the quadratic phase term to be $\sim$-7.4 $ps^2$ (equivalent to a single photon traversing $\sim$ 343 meters of SMF-28e fiber with $\beta_2=-2.16\times10^{-2}$ $ps^2/m$). This corresponds to the 313 m fiber spool plus an estimated $\sim$15 m of fiber path for each of signal and idler photons in the remainder of the apparatus.
We compensate for the total estimated dispersion by applying an equal amount but opposite sign of quadratic spectral phase onto the signal photon bins using Pulse shaper 1. The interferogram acquired after such compensation is overlayed onto Figs.~\ref{3Rep_RF}(c) and \ref{3Rep_RF}(d).
The normalized coincidences in Fig.~\ref{3Rep_RF}(c)(blue) are restored almost identically to the interferogram obtained in the absence of the spool. The peak coincidences improve by a factor of 2.3 
after dispersion compensation as shown in Fig.~\ref{3Rep_RF}(d), close to the theoretically expected factor of 2.2. 

Although here we compensate for dispersion prior to routing the photons to different paths, the dispersion can also be compensated anywhere down the fiber links prior to modulation.
Furthermore, in an effect known as nonlocal dispersion compensation \cite{fransonnonlocal,fitchfransondispersion}, the total dispersion accumulated by the biphoton can be compensated in only one of the links by controlling the spectral phase of either the signal or the idler photon.

Our results demonstrate the capability to achieve the original delay sensitivity of the interferogram despite a $\sim$1.5 $\mu s$ mismatch between the biphoton delays. This highlights the important nonlocal delay sensing capabilty of our approach. 







\begin{figure*}[!htb]
\centering
\includegraphics[width=\textwidth]{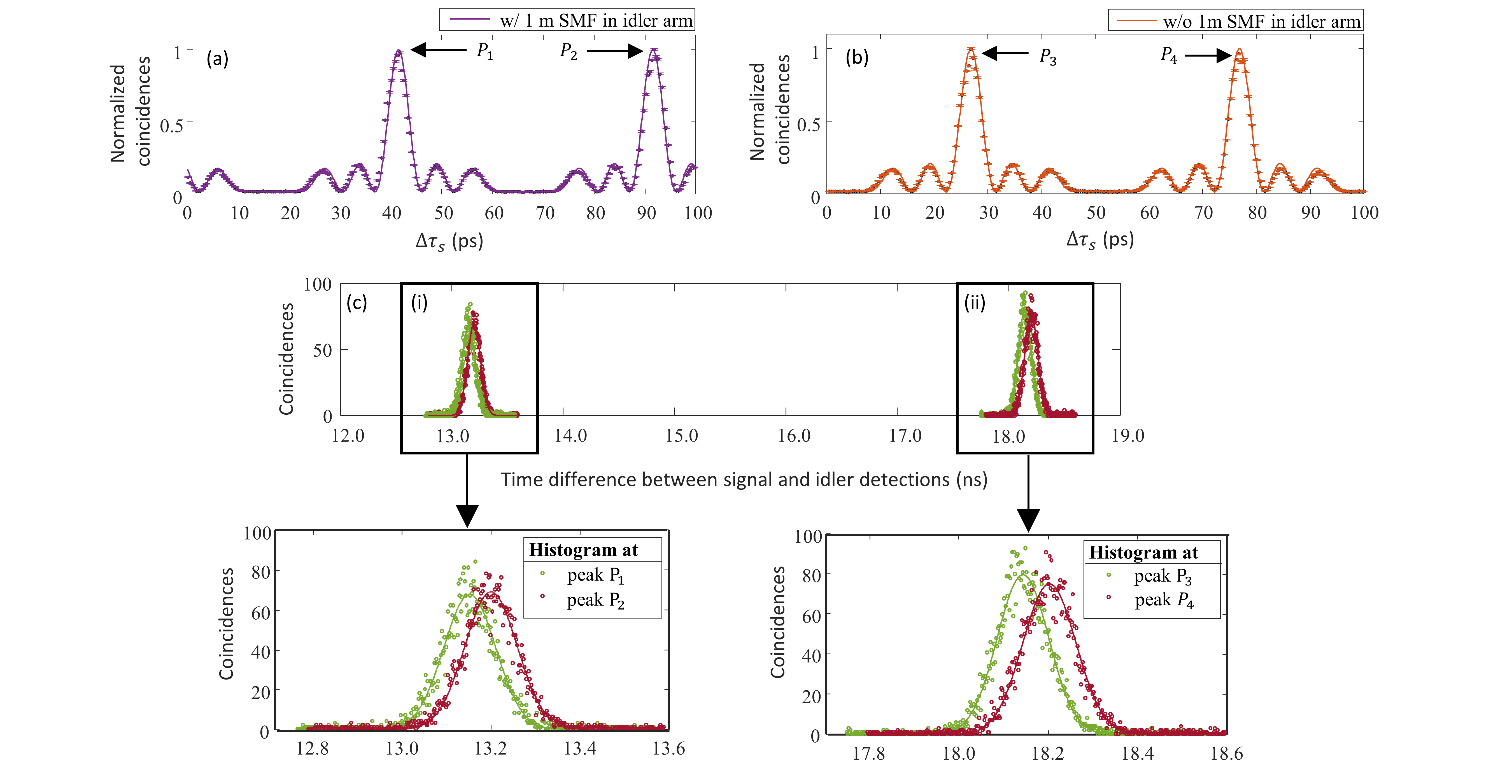} 
\caption{ (a)-(b) Two-photon coincidence interferograms recorded with and without an approximately 1 m-long SMF in the idler arm, respectively.  $\Delta\tau_s$ refers to the delay added into the signal arm via the motion of the ODL (scanned in steps of 0.4 ps), referenced to the same starting point ($\Delta\tau_s$ = 0) for both interferograms. Solid lines: Normalized theoretical coincidence probability fit to the experimental curves. (c) Coincidence time-tag histograms acquired when the ODL in the signal arm is set to the delay positions corresponding to the datapoints at peak $P_1$ and $P_2$ from (a), and the datapoints at peak $P_3$ and $P_4$ from (b). The time-bins in the histograms are 2 ps-wide. Solid lines: Gaussian fit to time-tag histograms.}
\label{Hist}
\end{figure*}

\subsection{Unambiguous sensing by complementing with detection time-tags}\label{UnambigiousSensing}
Although periodic modulation of biphotons allows for sensing
relative delays despite large imbalance in their path
lengths, it comes at a price: the interferogram as a function of relative delay repeats every modulation period and restricts the unambiguous sensing range. One can circumvent this limitation by supplementing the interferogram with photon arrival time-tags recorded by the event timer after detection. In this section, we demonstrate a basic implementation that utilizes histograms generated by the event timer to determine the coarse delay, thereby removing the ambiguity from the periodic interferogram. So long as the resolution of the detectors and timing electronics is less than the repetition period, unambiguous sensing of delay can be performed. We demonstrate this scheme by comparing measurements performed with and without an SMF (FS P/N: SM-FCU-FCU-SX-FS-1M-PVC) of length $1.05 \pm 0.05$ m in the idler arm of our setup prior to phase modulation (refer to Fig.~\ref{setup}).
In this experiment, the modulation frequency is set to $\omega_{_{RF}} = \omega_{_{FSR}} = $ 20 GHz to result in a repetition period ($T_{rep}$) of  $50$ ps. Note that the fiber inserted in the idler adds a delay (of $\sim$5 ns) which is $\sim$100 times larger than the RF modulation period.

We perform measurements with and without the additional fiber in the idler arm and adjust the ODL to position at an interferogram peak in each case---at which point time tagger-histograms of the
difference between the signal and idler detection times are acquired. Positive values of the delay difference signify that the signal is detected later than the idler. Then the difference between the mean values of the time tagger-histograms should be an integer multiple of $T_{rep}$, i.e., $kT_{rep}$ where $k \in \mathbb{Z}$. We thus rely on the time tagging electronics to disambiguate the value of the integer $k$. One can easily show that the effective delay that was inserted into the idler arm $\Delta \tau_{_I}$ should be equal to:
\begin{equation}\label{delay1mSMF}
\Delta \tau_{_I} = kT_{rep} + \Delta\tau^{(w/)}_{_S} - \Delta \tau^{(w/o)}_{_S}  
\end{equation} where $\Delta\tau^{(w/)}_{_S}$ and $\Delta \tau^{(w/o)}_{_S}$ are the delay settings (of the ODL) in the signal arm at the peak of the interferograms acquired with and without the additional SMF in the idler arm respectively.
The interferograms measured in the experiment are shown in Figs.~\ref{Hist}(a-b).  The time tagger-histograms acquired when the ODL is positioned at the interferogram peaks are shown in Figs.~\ref{Hist}(c). The peaks from Fig~\ref{Hist}(a) are denoted by $P_i$ and those from Fig~\ref{Hist}(b) by $P_j$, where $i \in \{1,2\}$ and $j \in \{3,4\}$.

Let $\tau_h(P_i)$ denote the mean value from the Gaussian fit of the histogram acquired at the interferogram peak $P_i$; the estimated means with the confidence bounds are tabulated in \textit{Supplement D}. In our experiment, the difference $\tau_h(P_4) - \tau_h(P_2) $ is obtained to be 5002 ps with a 95\% confidence interval-width $\lesssim 6$ ps. As previously stated, we expect this difference to be a integer multiple of the repetition period. Since the estimate of $\tau_h(P_4) - \tau_h(P_2) $ is localized much tighter than $T_{rep}$, it can be rounded to 100 $T_{rep}$, i.e., the integer $k{_{_{P_4-P_1}}} = 100$. 

Further, the difference between the ODL settings at the interferogram peaks $P_i$ and $P_j$ denoted by $\Delta\tau^{(w)}_{_S}(P_i) - \Delta\tau^{(w/o)}_{_S}(P_j)$ are estimated from a weighted least squares fit of the interferograms to the theory and listed in \textit{Supplement D}. For instance, $\Delta\tau^{(w/)}_{_S}(P_2) - \Delta\tau^{(w/o)}_{_S}(P_4) = $14.65 $\pm$ 0.04 ps; the error bars denoting the 95\% confidence bounds. The effective delay due to the $\sim$1 m SMF in the idler arm computed from Eq.~(\ref{delay1mSMF}) is $5014.65 \pm 0.04$ ps, consistent with the manufacturer specifications. The result is verified for different pairs of interferogram peaks as shown in \textit{Supplement D}.

\subsection{Relation to second order time-correlation function}


\begin{figure}[!htb]
\centering
{\includegraphics[width=\linewidth]{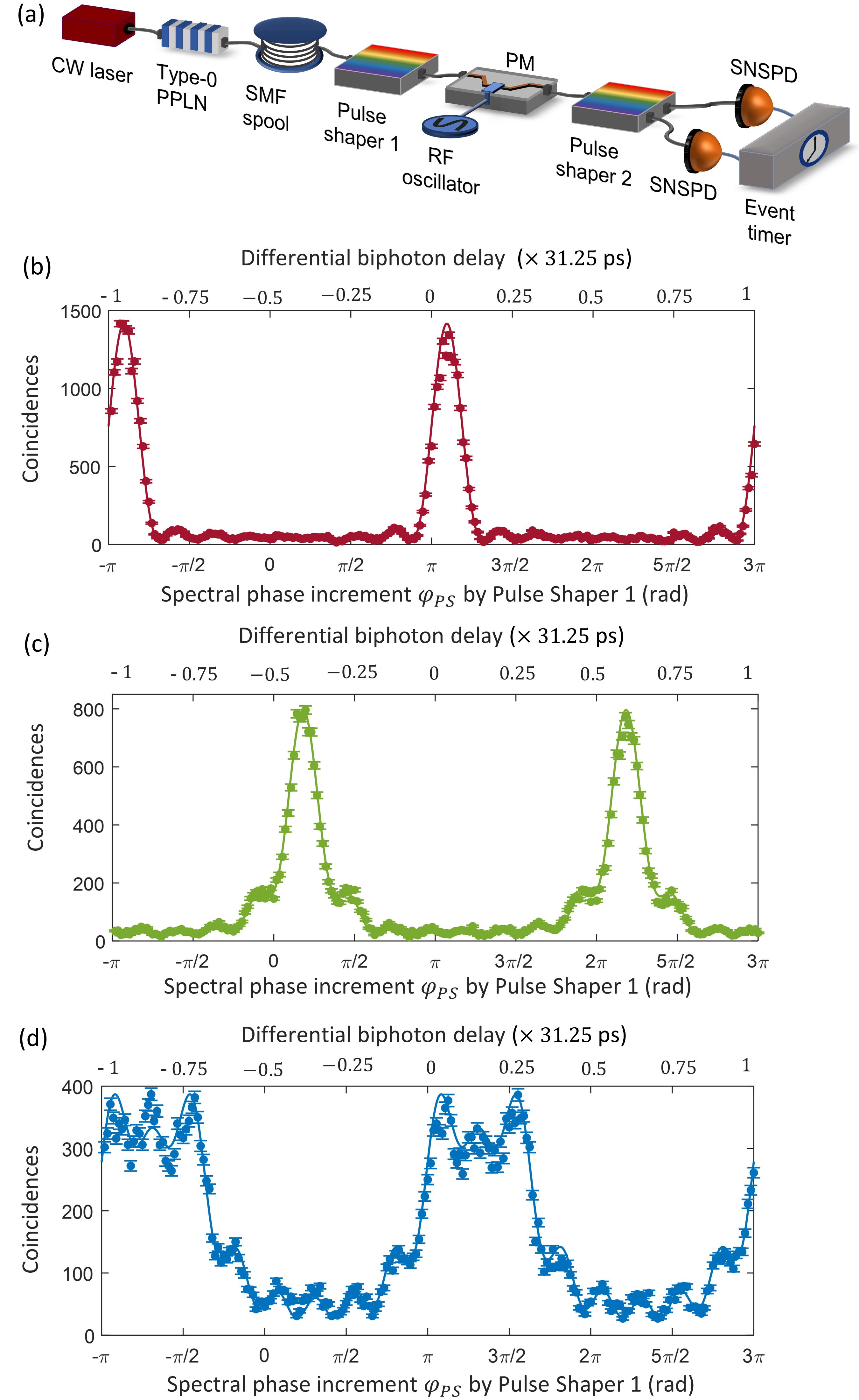}}
\caption{(a) Experimental setup in local detection geometry.
Two-photon coincidence-interferograms measured after equalized-amplitude mixing of frequency bins onto the bin pair selected for detection,  recorded as slope of linear spectral phase on the signal bins is scanned, (b) without additional fiber in the signal arm, (c) with a 103 m-long SMF spool inserted in the signal arm, (d) with a 210 m-long SMF spool inserted in the signal arm. Solid lines: Theoretical fit. }
\label{TimeCorr}
\end{figure}

Our experiments based on the coincidence measurements after frequency mixing are closely related to Glauber's second order time-correlation function. Consider the BFC described in Eq.~(\ref{SS1}) with frequency bins with infinitely narrow lineshape. The Glauber's second order time-correlation function of such a two-photon state is given by

\begin{equation}\label{SS8}
G^{(2)}(t_{_S},t_{_I}) = \left| \sum_{k=-N}^{N} \alpha_k e^{-i k\omega_{_{FSR}}(t_{_S}-t_{_I})}  \right|^2,
\end{equation} where $t_{_S}$ and $t_{_I}$ refer to the arrival times of the signal and idler at the SPDs. Setting $\tau_{_S}$ and $\tau_{_I}$ as delays in the signal and idler paths such that  $t_{_S} = t -\tau_{_S}$ and $t_{_I} = t -\tau_{_I}$, the second order correlation is equivalent to that of Eq.~(\ref{SS7}) when the spectral mixing amplitudes are set to be
equal i.e $C_k = 1$ $\forall$ $k$.



Our experimental setup is shown in Fig.~\ref{TimeCorr}(a).  We insert different lengths of SMF-28e fiber after the SPDC source to introduce dispersive reshaping of the biphoton time-correlation function. For simplicity we use a local detection geometry: a common EOPM modulates the biphotons at a frequency of 32 GHz and a modulation depth $m = 4.48$ rad.  Pulse shaper 1 performs two functions: (1) it imparts a linear spectral phase onto the signal frequency bins in order to vary the signal-idler delay; and (2) it imposes a frequency-dependent loss which compensates for the variation in the Bessel frequency mixing amplitudes.  In particular, we program the pulse shaper such that the relative electric field amplitudes of  signal and idler frequency bins are each scaled according to $|J_k(m)|^{-1}$.  This procedure equalizes the $|C_k|$ in Eq.~(\ref{SS7}), ensuring that our measurement provides the time correlation function. Pulse shaper 2 finally selects 15 GHz-wide frequency bins from centers of signal and idler spectra and routes them to different SNSPDs for coincidence detection.
Fig.~\ref{TimeCorr}(b) corresponds to the interferogram acquired without additional fiber; the interferograms in Figs.~\ref{TimeCorr}(c) and \ref{TimeCorr}(d) are measured with 103 m and 210 m of additional fiber inserted, respectively. The data reveal a modest broadening of the time correlation function in Fig.~\ref{TimeCorr}(c) and a more pronounced broadening in Fig.~\ref{TimeCorr}(d), corresponding to the increased length of dispersive fiber.


According to Eq.~(\ref{EqDispNL}), with equal values for the mixing coefficients $C_k$ and with equal fiber lengths for signal and idler ($L_S = L_I = L$), the coincidence probability---and the second-order time corrleation function---is given by:
\begin{equation}\label{SS9}
\begin{aligned}
&\mathcal{P}(\tau) \propto \Bigg| \sum_{k=-N}^{ N}  \exp {\Big[ i\beta_2 Lk^2 \omega^2_{_{FSR}}  +ik\omega_{_{FSR}}\tau_{eff}\Big]}  \Bigg|^2,\\
&\tau_{eff} = (\tau + 2\beta_2 L \Omega_0) +\omega_{_{FSR}}^{-1}[\varphi_{_{PS}}+\pi],
\end{aligned}
\end{equation} Here $\beta_2$ is the fiber dispersion, and  
Eq.~(\ref{SS9}) indicates that dispersion induces both quadratic and linear spectral phases proportional to $\beta_2 L$. Thus, both broadening and horizontal translation of the interferograms are expected.
The experimental interferograms in Figs.~\ref{TimeCorr}(b,c,d) are fit with the theory for SMF-28e fiber lengths of 9 m, 112 m and 219 m respectively.  These values correspond to the lengths stated above, with an additional 9 m added to account for residual dispersion in the  remainder of our setup.  Both the shifts of the interferograms along the delay axis and their broadening and reshaping closely match the theoretical curves.  These results clearly link our measurement technique to the biphoton time correlation function---showing that in addition to the subpicosecond delay sensitivity emphasized throughout this paper, characterization of temporal reshaping of the biphoton is also accessible.







\section{ Discussion and conclusion}\label{discussion}

In summary, we present a nonlocal sensing scheme with subpicosecond delay resolution requiring only off-the-shelf telecom equipment and resources expected to be staples of quantum network infrastructure. By interfering multiple frequency bins in the SPDC spectrum, we map the spectral phase accumulated by a photon traversing through a device under test to a change in the probability of coincident events between that photon and its entangled counterpart. This is accomplished via phase modulation followed by selective spectral filtering and coincidence detection. The proposed approach based on spectral quantum interferometry does not rely on spatial overlap of biphotons---thus sensing in a nonlocal architecture is feasible. 

Our experiments demonstrate capability for sensing RF phase shifts as well as relative biphoton delays with errors (representing 95\% confidence widths) on the order of $\pm$$0.7^{\circ}$ and $\pm$$0.04$ ps, respectively.
Supplemented by dispersion compensation, our approach handles microsecond-scale delay differences between the optical links traversed by biphotons. 
Although the unambiguous range from measurements solely based on the periodic interferogram is restricted to inverse of the modulation frequency, we have shown that detection time-tags from SPDs can be used to resolve this ambiguity in coarse delay.

 This work helps to elucidate the connection between phase modulation based methods \cite{olislager2010frequency, kues2017chip, imany201850, imany2018characterization} for demonstrating frequency bin entanglement and the biphoton time correlation function. This capability may potentially be exploited for quantum state tomography of two-photon time-frequency entangled states.


Note that distance measurements using classical dual frequency combs with slightly different repetition rate is know to offer long range sensing as well as high resolution \cite{zhu2018dual,trocha2018ultrafast}. Analogously, by applying two closely spaced RF frequencies to the modulator in our entangled photon measurements, one can increase the interferogram repetition period to the  inverse of the RF frequency separation and thereby realize a large nonambiguous range.

Our scheme can potentially be investigated further to perform nonlocal clock synchronization \cite{giovannetti2001clock} or to use the quantum signals for both time transfer and network protocols such as secret-key generation \cite{dai2020towards}.  While the demux and filtering operations are performed using pulse shapers in our demonstration, they can instead be realized with only coarse and dense wavelength division multiplexers. Thus, the proposed scheme requires only off-the-shelf telecom equipment.

\section*{Acknowledgments}

The authors thank AdvR for loaning a PPLN ridge waveguide and Dr.~Joseph M. Lukens for valuable discussions. The funding for this work was provided by the National Science Foundation (1839191-ECCS and 1747426-DMR).

\section*{Supplement}
\label{methods}

\subsection{Theory}

 Here, we describe the theory to obtain probability of coincidence attuned to our experimental realization in the main text. We generate entangled photon pairs through spontaneous parametric down-conversion of a continuous-wave laser at frequency $2\omega_0$. Correlated frequency slices (with a bandshape function $\mathcal{F}$) offset from the center of the SPDC spectrum by $\Omega_0$, 
each with a bandwidth of $\Delta\Omega$ are selected from the signal and idler spectra. 
Consider the input state,
\begin{equation}\label{A1}\begin{aligned}
\ket{\Psi'}  = \int_0^\infty d\Omega\, &\Phi(\Omega) \mathcal{F}(\Omega-\Omega_0)\\ &[\hat{a}^{(S)}(\omega_0+\Omega)]^\dagger [\hat{a}^{(I)}(\omega_0-\Omega) ]^\dagger \ket{\mathrm{vac}},
\end{aligned}
\end{equation} where $\ket{\mathrm{vac}}$ is the vacuum state, and $\Phi(\Omega)$ is the broadband phase matching function of the SPDC process.

The input electric field operators are given by

\begin{equation}
\begin{aligned}
\label{EfieldSAA}
\hat{E}_{S,in}^{(+)}(t) \propto \int &d\omega_{_S}\, e^{-i\omega_{_S} t}\hat{a}^{(S)}(\omega_{_S}),
\end{aligned}
\end{equation}

\begin{equation}
\begin{aligned}
\label{EfieldIAA}
\hat{E}_{I,in}^{(+)}(t) \propto \int &d\omega_{_I}\, e^{-i\omega_{_I} t}\hat{a}^{(I)}(\omega_{_I}).
\end{aligned}
\end{equation}

The signal and idler spectral slices are routed to different fiber links where they traverse through delays $\tau_{_S}$ and $\tau_{_I}$, and pick up additional complex spectral amplitudes $ H^{(S)}$ and $ H^{(I)}$ respectively. The annihilation operators $\hat{a}^{(S)}$  and $\hat{a}^{(I)}$ corresponding to the signal and idler transform into $\hat{b}^{(S)}$ and $\hat{b}^{(I)}$ after traversing down the link (prior to phase modulation) as follows:

\begin{equation}\label{operators_S}
  \hat{b}^{(S)}(\omega_{_{S}}) = \hat{a}^{(S)}(\omega_{_{S}}) e^{i\omega_{_{S}}\tau_{_S}} H^{(S)}(\omega_{_S}),
\end{equation}

\begin{equation}\label{operators_I}
  \hat{b}^{(I)}(\omega_{_{I}}) = \hat{a}^{(I)}(\omega_{_{I}}) e^{i\omega_{_{I}}\tau_{_I}} H^{(I)}(\omega_{_I}).
\end{equation}

 Down the links, we consider phase modulation of the following form applied on the signal and idler respectively.
 
 \begin{equation}\begin{aligned}\label{modTimeS}
m_S(t) &= \exp{(-im \sin(\omega_{_{FSR}} t + \phi_{_{S}}))}\\
&= \sum_{k=-\infty}^\infty J_{k}(m) e^{-ik\phi_{_S} - ik\omega_{_{FSR}}t},
 \end{aligned}
 \end{equation}
 
 \begin{equation}\begin{aligned}\label{modTimeI}
m_I(t) &= \exp{(-im \sin(\omega_{_{FSR}} t + \phi_{_{I}}))}\\
&=\sum_{k=-\infty}^\infty J_{k}(m) e^{-ik\phi_{_I} - ik\omega_{_{FSR}}t}.
 \end{aligned}
 \end{equation}
 
 

After the phase modulation, the signal and idler annihilation operators transform as:

\begin{equation}\label{phaseMod_S}
\hat{c}^{(S)}(\omega_{_S})= \sum_{k=-\infty}^\infty J_{k}(m) e^{-ik\phi_{_S}}\hat{b}^{(S)}(\omega_{_S} - k\omega_{_{FSR}}),
\end{equation}

\begin{equation}\label{phaseMod_I}
\hat{c}^{(I)}(\omega_{_I})= \sum_{k=-\infty}^\infty J_{k}(m) e^{-ik\phi_{_I}}\hat{b}^{(I)}(\omega_{_I} - k\omega_{_{FSR}}).
\end{equation}
 
 Following this, filters with lineshape $f(\omega)$ of spectral width $\delta\Omega$ and symmetric passbands at the center of signal and idler spectra i.e., at $\omega_0 \pm \Omega_0$, are used to select correlated bin pairs. 
 
 The output electric field operators are given by

\begin{equation}
\begin{aligned}
\label{EfieldSAA_out}
\hat{E}_{S,out}^{(+)}(t) \propto \int &d\omega_{_S}\, f(\omega_{_S}- (\omega_0+\Omega_0)) \hat{c}^{(S)}(\omega_{_S}) e^{-i\omega_{_S} t},
\end{aligned}
\end{equation}

\begin{equation}
\begin{aligned}
\label{EfieldIAA_out}
\hat{E}_{I,out}^{(+)}(t) \propto \int &d\omega_{_I}\, f(\omega_{_I}- (\omega_0-\Omega_0))\hat{c}^{(I)}(\omega_{_I}) e^{-i\omega_{_I} t}.  \end{aligned}
\end{equation} Here the modulation frequency is chosen with respect to the total spectral bandwidth such that $\Delta\Omega/\omega_{_{RF}}$ is set to  $2N+1$, representing the total frequency bin pairs accommodated in the signal and idler bandwidth. The width of the filter $f(\omega)$ is set to be well within the modulation frequency, i.e., $\delta\Omega \ll \omega_{_{RF}}$.

The probability of coincidence detection (normalized over the acquisition window $\Delta t$) between the selected signal-idler central frequency-bins at the output is given by:

\begin{equation}
\label{SS_CoinProp}\begin{aligned}%
{\mathcal{P}}&(\tau,\phi_{_{RF}}) \propto \\  & \frac{1}{\Delta t}\int_{\Delta t} dt \int_{\Delta T} dT\, \left|\left\langle\mathrm{vac} \bigg| \hat{E}_{S,out}^{(+)}(t+T) \hat{E}_{I,out}^{(+)}(t) \bigg| \Psi' \right\rangle  \right|^2,
\end{aligned}
\end{equation} where the integral over $T$ extends over $\Delta T$, taken to be longer than the wavepacket duration. Inserting Eqs.~(\ref{A1}-\ref{EfieldIAA_out}) into Eq.~(\ref{SS_CoinProp}), and assuming the biphoton joint spectral amplitude (JSA), consisting of phase matching function $\Phi(\Omega)$ and the complex amplitude functions $H^{(S)}$ and $H^{(I)}$, to be slowly varying with respect to the narrow width $\delta\Omega$ of the frequency filter $f$, allows us to effectively consider the JSA at the center frequencies of the signal-idler filters $\omega_0 \pm \Omega_0$. That is,

\begin{equation}\label{A2}\begin{aligned}
{\mathcal{P}}(\tau,\phi_{_{RF}}) &\propto \int  d\Omega\,\Big|f(\Omega - \Omega_0)f(-(\Omega - \Omega_0)) \chi(\Omega,\tau,\phi_{_{RF}})\Big|^2\\
&\appropto \left|\chi(\Omega_0,\tau,\phi_{_{RF}})\right|^2,
\end{aligned}
\end{equation} where $\chi(\Omega_0,\tau, \phi_{_{RF}})$ describes the probability amplitude of the selected central frequency bin pair with sideband contributions from the rest of the bins as given below:

\begin{equation}\label{A3}
\begin{aligned}
\chi(\Omega_0,\tau, \phi_{_{RF}}) =&\sum_{k=-N}^{N} C_k  e^{i k(\omega_{_{FSR}}\tau + \phi_{_{RF}})} \Big[\Phi(\Omega_0 + k\omega_{_{FSR}}) \\
&\mspace{10mu}H_{_S}(\omega_0 +\Omega_0 + k\omega_{_{FSR}}) H_{_I}(\omega_0 -\Omega_0 - k\omega_{_{FSR}})\Big],\\
\end{aligned}
\end{equation}

\begin{equation}\label{Params}
\begin{aligned}
&\tau = \tau_{_S} - \tau_{_I},\\
&\phi_{_{RF}} = \phi_{_S} - \phi_{_I}.
\end{aligned}
\end{equation} The mixing coefficient \mbox{$C_k$ $= J_k(m)J_{-k}(m)$  $ = |J_k(m)|^2e^{ik\pi}$}, where $J_{-k}(m)$ and $J_{k}(m)$ are Bessel functions of the first kind.  Here, we have considered nearly rectangular spectral slices from the signal and idler spectrum given by the bandshape function $\mathcal{F}$ accommodating 2N+1 frequency bins.

 Effectively as a result of phase modulation and subsequent spectral filtering operations, the measured coincidence probability corresponds to a post-selected-BFC with 2N+1 dimensions and a free spectral range ($\omega_{_{FSR}}$) equal to the modulation frequency $\omega_{_{RF}}$.



 The differential delay between the biphotons denoted by $\tau$, the relative phase between RF modulating sinusoids $\phi_{_{RF}}$, and the complex spectral amplitudes $H^{(S)}$ and $H^{(I)}$ constitute the control parameters of this scheme defining the interferogram. Note that even when the detector resolution $T_R$ is much larger than the wavepacket duration and the integral over $T$ in Eq.~(\ref{SS_CoinProp}) extends over $\Delta T \sim T_R$, the resultant interferogram from the above described scheme can still distinguish changes to differential biphoton delay $\Delta\tau \ll T_R$.   Note that the individual delays ($\tau_{_S}$ and $\tau_{_I}$) and the RF phases ($\phi_{_S}$ and $\phi_{_I}$) are both defined relative to a common overall clock.  
 

In the experiment we utilize a pulse shaper to route the biphotons to different paths; the pulse shaper is also used in some of our demonstrations to impart linear spectral phase on the frequency bins \cite{weiner2011ultrafast}. We let $k\varphi_{_S}$ and $k\varphi_{_I}$ be the spectral phases imparted by the pulse shaper on the $k^{\mathrm{th}}$ signal and $k^{\mathrm{th}}$ idler bin respectively. We define the complex amplitudes at $k^{\mathrm{th}}$ signal and $-k^{\mathrm{th}}$ idler bin as

\begin{equation}\begin{aligned}
&H_{_S}(\omega_0 +\Omega_0 + k\omega_{_{FSR}}) = e^{i{k\varphi_{_S}}},\\ 
&H_{_I}(\omega_0 -\Omega_0 - k\omega_{_{FSR}}) = e^{-ik\varphi_{_I}}.
\end{aligned}\label{HsHi1}
\end{equation}

Additionally we set

\begin{equation}\begin{aligned}\label{alphak}
\alpha_k = \Phi(\Omega_0 + k\omega_{_{FSR}}).
\end{aligned}
\end{equation}

The coincidence probability can then be rewritten as

\begin{equation}\label{A6}\begin{aligned}
    \mathcal{P}(\Delta\phi) \propto& \left| \sum_{k=-N}^{N}  \alpha_k C_{k}  e^{i k(\Delta\phi)}  \right|^2
\end{aligned}
\end{equation}

where, 
\begin{equation}\label{A7}
\begin{aligned}
&\Delta\phi = \omega_{_{FSR}}\tau + \phi_{_{RF}} + \varphi_{_{PS}}
&\tau = \tau_{_S} - \tau_{_I}&\\
&\phi_{_{RF}} = \phi_{_S} - \phi_{_I} 
&\varphi_{_{PS}} = \varphi_{_S} - \varphi_{_I}\\
\end{aligned}
\end{equation}

We can infer that the coincidence probability repeats every $2\pi$ rad as a function of the effective spectral phase increment $\Delta\phi$. 

Note that when $\varphi_{_S}$ and $\varphi_{_I}$ are changed in common mode (so $\varphi_{_I}$ = $\varphi_{_S}$, the interferogram is unchanged.  In the terminology defined here, the case $\varphi_{_I}$ = $\varphi_{_S}$ corresponds to identical delays applied to signal and idler photons; their relative delay is unaltered.  For the biphoton states studies here, our measurement scheme is insensitive to common-mode delays of signal and idler photons. The measurement is only sensitive to changes in their relative delay. Differential delay between signal and idler photons can be obtained by programming the pulse shaper \cite{pe2005temporal,lukens2013biphoton} such that it imposes opposite linear spectral phase on the signal and idler ($\varphi_{_I}$ = - $\varphi_{_S}$).

 \subsubsection{Coincidence probability in the presence of dispersion}

Consider the complex amplitude functions $H^{(S)}$ and $H^{(I)}$ in Eq.~(\ref{HsHi1}) to include the chromatic dispersion in signal and idler arms respectively.

\begin{equation}\label{A11}\
\begin{aligned}
&H^{(S)}(\omega_0 + \Omega_0 + k\omega_{_{FSR}} )= e^{ \Big[\frac{1}{2} i \beta_2 ( \Omega_0 + k\omega_{_{FSR}})^2 L_s +i{k\varphi_{_S}}\Big]}\\ 
  &H^{(I)}(\omega_0 - \Omega_0 - k\omega_{_{FSR}} )= e^{ \Big[\frac{1}{2} i \beta_2 (\Omega_0 + k\omega_{_{FSR}})^2 L_I -i{k\varphi_{_I}}\Big]}
\end{aligned}
\end{equation} where $\beta_2 $ is the dispersion parameter of SMF-28e, $L_S$ and $L_I$ are effective lengths of SMF-28e corresponding to the dispersion accumulated by signal and idler respectively.

Considering nearly constant phase matching function $\Phi(\Omega)$ across 2N+1 frequency bins, the coincidence probability from Eq.~(\ref{A6}) in the presence of dispersion in the signal and idler arms takes the form shown in 
Eq.~(\ref{EqDispNL}) in the main text.

\begin{table*}[htbp]
\centering
\begin{tabular}{ccc}
\hline
Figures   \hspace{2mm}  &$\Delta t$  \hspace{6mm} $\Delta T$ \hspace{8mm} $\Delta \Omega$ \hspace{10mm} $\Omega_0$ \hspace{9mm} $\omega_{_{FSR}}$ \hspace{8mm} $\delta \Omega$ \hspace{2mm}\\
\hline

Fig. \ref{3Rep_RF}(a),(b),(c: green \& orange markers) \hspace{3mm}  &\hspace{3mm}5 s \hspace{3mm}256 ps  \hspace{3mm} 302 GHz \hspace{3mm} 608 GHz \hspace{3mm} 32 GHz \hspace{3mm} 15 GHz\\

Fig. \ref{3Rep_RF}(c: blue marker)(d), Fig. \ref{TimeCorr}  \hspace{3mm} &\hspace{1.5mm}10 s \hspace{3mm}256 ps \hspace{3mm} 302 GHz \hspace{3mm} 608 GHz \hspace{3mm} 32 GHz \hspace{3mm} 15 GHz\\

Fig. \ref{Hist}   \hspace{3mm}  &\hspace{3mm}5 s  \hspace{3mm}256 ps \hspace{3mm} 190 GHz \hspace{3mm} 200 GHz \hspace{3mm} 20 GHz \hspace{3mm} 11 GHz\\ 
\hline
\end{tabular}

\caption{Total coincidence-acquisition time ($\Delta t$),  the histogram-time window ($\Delta T$) over which coincidences are integrated to plot the datapoints in the figures, carved spectral width ($\Delta \Omega$) and offset from the SPDC center ($\Omega_0$), modulation frequency ($\omega_{_{FSR}}$) and the frequency bin width ($\delta\Omega$). }
  \label{ExptParams}
\end{table*}

\begin{table*}[htbp]
\centering
\begin{tabular}{ccc}
\hline
Coefficients(95\% confidence bounds) &\hspace{9mm}$A$ in ps \hspace{25mm} $B$ in ps \hspace{20mm} $C$ in a.u. \hspace{9mm} \\

\hline

Histogram at $P_1$  &\hspace{0mm} 13152 (13150,13154) \hspace{3mm} 82.80 (81.06,84.52) \hspace{8mm}67.64 (66.41,68.86)  \\

Histogram at $P_2$  &\hspace{0mm} 13200 (13198,13200) \hspace{3mm} 83.18 (81.70,84.68) \hspace{8mm}68.68 (67.91,70.05)  \\

Histogram at $P_3$  &\hspace{0mm} 18144 (18142,18144) \hspace{3mm} 81.32 (79.78,82.86) \hspace{8mm}79.92 (78.62,81.23)  \\

Histogram at $P_4$  &  \hspace{0mm} 18202 (18202,18204) \hspace{3mm} 84.56 (83.00,86.12) \hspace{8mm}75.44 (74.24,76.65)\\

\hline
\end{tabular}

\caption{Estimated coefficients and the 95\% confidence bounds from the least squares fit of Histograms in Fig.\ref{Hist}(c) using a Gaussian model $y = f(x) = Ce^{-(\frac{x-A}{B})^2}$.}
  \label{Gaussianfit}
\end{table*}

\subsection{Choice of modulation index}

\begin{figure}[ht!]
\centering
{\includegraphics[width=\linewidth]{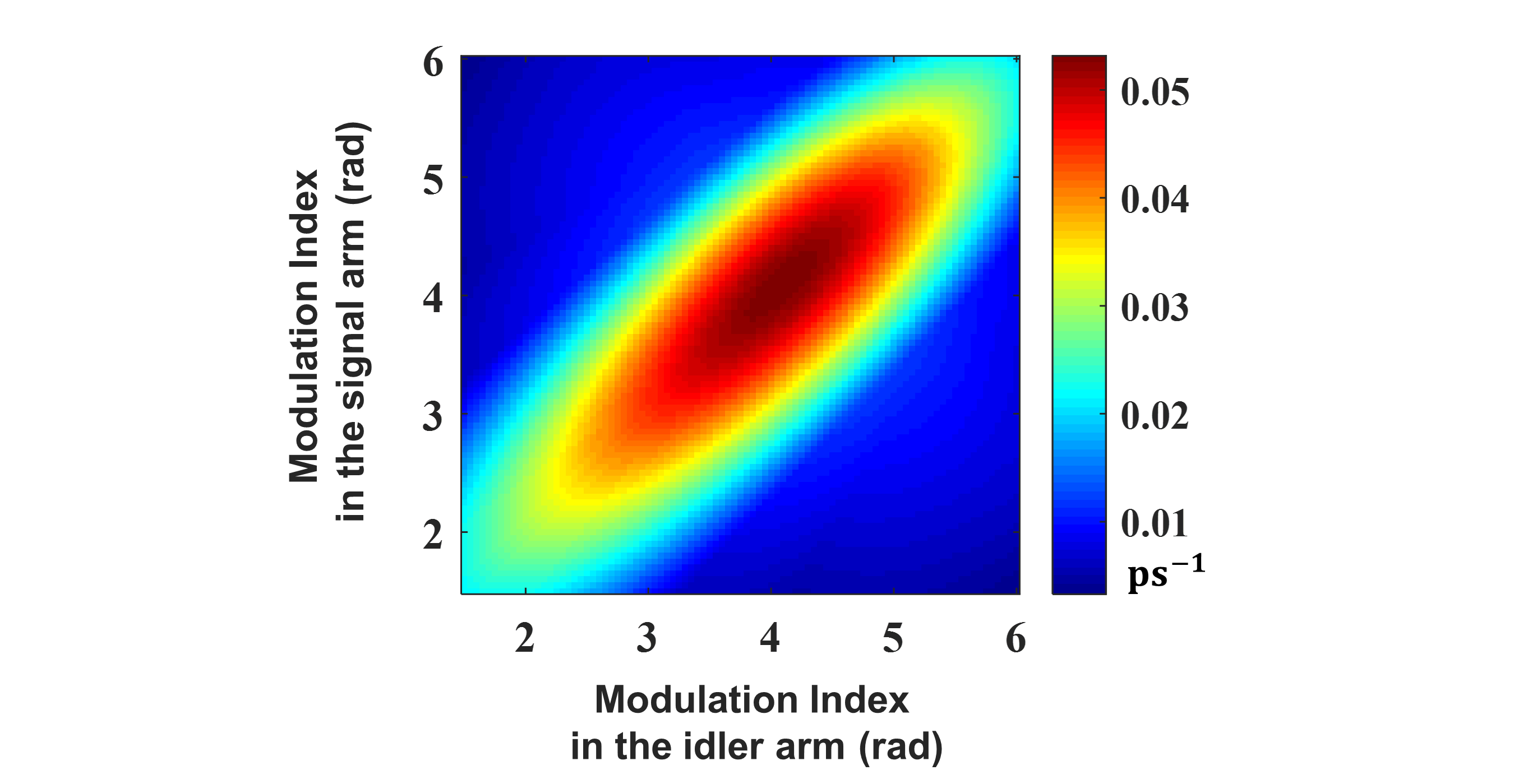}}
\caption{ Maximum slope in theoretical coincidence probability curves as a function of differential biphoton delay at each pair of modulation depth settings applied onto the signal and idler photons.}
\label{Supp_Slope}
\end{figure}

In the proposed sensing scheme, signal and idler are routed to different optical links where they are phase modulated to allow for coherent frequency bin mixing. We have shown that after phase modulation, coincidences measured between central frequency bins from the signal and idler spectrum are sensitive to the differential delay traversed by the two photons. The depth of sinusoidal phase modulation applied to the biphotons determines the amplitudes of the sidebands that are involved in the mixing of frequency bins. The choice of modulation thus dictates the sensing resolution by altering (1) the width of the coincidence curves as a function of differential biphoton delay, (2) the peak coincidence value controlled by the sideband power lost outside the computational space (nine signal and nine idler frequency bins).  We use the maximum slope in the theoretical coincidence probability with respect to differential biphoton delay as a metric that takes both the above-described-effects into account for determining optimal modulation index settings. As in Fig.~\ref{Theory_fig}(c) of the main text, we consider the coincidence probability curves without normalization, as that includes the effect of loss that occurs due to sideband spreading outside the computational space.  In Fig.~\ref{Supp_Slope} we show the maximum slope of the theoretical coincidence probability curves for various modulation depths applied on to the signal and idler photons, respectively, for the case of a nine dimensional BFC with FSR of 32 GHz.  The higher the maximum slope metric, the better the sensing resolution. The slope is plotted in units of  $ps^{-1}$ as the theoretical curves are functions of delay. We observe that the maximum slope is optimized when the signal and idler are modulated at equal depths close to \mbox{$\sim$ 4.1 radians}. In our experiments we set the modulation depth of both the photons at 4.48 radians lying within a 4\% deviation contour from the optimal operating point.

\subsection{Experimental details}

The details on the experimental parameters across different sections are noted in Table~\ref{ExptParams}.

The key devices employed in our experiments with their manufacturers and model number are as follows : 

\begin{itemize}
    \item  Periodically poled lithium niobate waveguide: AdvR.
    \item Pulse Shaper 1: Finisar WaveShaper 4000S, spectral resolution $\sim$ 18 GHz.
    \item Pulse Shaper 2: Finisar WaveShaper 1000S, spectral resolution $\sim$ 10 GHz. 
    \item Pulse Shaper 3: Finisar Waveshaper 1000SP, spectral resolution $\sim$ 10 GHz.
    \item Optical delay line: OZ Optics ODL-650-MC. 
    \item Electro-optic phase modulator: {EOSPACE 40 Gbps}.
    \item Superconducting nanowire single photon detectors: Opus One, Quantum Opus. 
    \item Event timer: HydraHarp 400, PicoQuant GmbH.
\end{itemize}

\subsection{Sensing delay by complementing the detection time-tags with interferogram}

A Gaussian model of the form $y = f(x) = Ce^{-(\frac{x-A}{B})^2}$ is used for the least squares fit of the time tagger-histograms in Figs.~\ref{Hist}(c), where $x$ is the delay in picoseconds and $y$ is the coincidences per histogram bin.  The estimated coefficients $\{A,B,C\}$ with the 95\% confidence bounds from the fit are given in Table~\ref{Gaussianfit}.

\begin{table}[htbp]
\centering
\vspace{1mm}
\begin{tabular}{ccc}

\hline
$\tau_h(P_4) - \tau_h(P_1)$ & \hspace{5mm}$5050$ ps $\approx 101T_{rep} = k_{_{P_4-P_1}}T_{rep}$  \\
$\tau_h(P_4) - \tau_h(P_2)$ & \hspace{5mm}$5002$ ps $\approx 100T_{rep} = k_{_{P_4-P_2}}T_{rep}$ \\
$\tau_h(P_3) - \tau_h(P_1)$ & \hspace{5mm}$4992$ ps $\approx 100T_{rep} = k_{_{P_3-P_1}}T_{rep}$  \\
$\tau_h(P_3) - \tau_h(P_2)$ &\hspace{5.5mm}$4944$ ps $\approx 99T_{rep}$ \hspace{0.5mm} $=  k_{_{P_3-P_2}}T_{rep}$  \\
\hline
\end{tabular}
\caption{ Difference between the mean values of histograms in Fig.\ref{Hist}(c). }
  \label{Coarse}
\end{table}

The coefficient A that represents the mean of the histogram acquired at a peak $P_i$ is denoted as $\tau_h(P_i)$. The difference between the histogram mean values is shown in Table~\ref{Coarse}. The difference between shifts in the peaks with respect to the ODL delay setting is estimated from a weighted least square fit of the theory to the interferograms in Fig.\ref{Hist}(a-b) and shown in Table~\ref{Fine}.

The 95\% confidence widths from the weighted least square fit of the interferograms in Figs.~\ref{Hist}(a) and \ref{Hist}(b) are 0.0686 ps and 0.0490 ps respectively. The resolution of the ODL given by the manufacturer specifications is 0.0017 ps (representing the standard error i.e., the width of 38\% confidence interval). The effective 95\% confidence width in our estimation is computed to be 0.0848 ps (i.e.,  $\sqrt{0.0686^2 + 0.0490^2 + 2(0.0017\times3.92)^2}$). The effective delay due to additional fiber inserted in the idler arm is given by Eq.~(\ref{delay1mSMF}), from the main text, and is consistent across all the values in Tables \ref{Coarse} and \ref{Fine}, i.e., $\big[k_{_{P_j-P_i}}T_{rep}\big] + \big[\Delta\tau^{(w/)}_{_S}(P_i) - \Delta \tau^{(w/o)}_{_S}(P_j)\big]$ = 5014.65 $\pm$ 0.04 ps, where $i \in \{1,2\}$ and $j \in \{3,4\}$.  




\begin{table}[htbp]
\centering

\vspace{1mm}
\begin{tabular}{ccc}

\hline
$\Delta\tau_{_S}(P_1) - \Delta\tau_{_S}(P_4)$ &  \hspace{5mm}$-35.35$ ps  \\
$\Delta\tau_{_S}(P_2) - \Delta\tau_{_S}(P_4)$ & \hspace{8mm}$14.65$ ps \\
$\Delta\tau_{_S}(P_1) - \Delta\tau_{_S}(P_3)$ & \hspace{8mm}$14.65$ ps\\
$\Delta\tau_{_S}(P_2) - \Delta\tau_{_S}(P_3)$ & \hspace{8mm}$64.65$ ps  \\
\hline
\end{tabular}
\caption{ The difference between shifts in the peaks estimated from a weighted least square fit of the theory to the interferograms in Fig.\ref{Hist}(a-b).}
  \label{Fine}
\end{table}

\bibliography{refs}

\begin{thebibliography}{43}%
\makeatletter
\providecommand \@ifxundefined [1]{%
 \@ifx{#1\undefined}
}%
\providecommand \@ifnum [1]{%
 \ifnum #1\expandafter \@firstoftwo
 \else \expandafter \@secondoftwo
 \fi
}%
\providecommand \@ifx [1]{%
 \ifx #1\expandafter \@firstoftwo
 \else \expandafter \@secondoftwo
 \fi
}%
\providecommand \natexlab [1]{#1}%
\providecommand \enquote  [1]{``#1''}%
\providecommand \bibnamefont  [1]{#1}%
\providecommand \bibfnamefont [1]{#1}%
\providecommand \citenamefont [1]{#1}%
\providecommand \href@noop [0]{\@secondoftwo}%
\providecommand \href [0]{\begingroup \@sanitize@url \@href}%
\providecommand \@href[1]{\@@startlink{#1}\@@href}%
\providecommand \@@href[1]{\endgroup#1\@@endlink}%
\providecommand \@sanitize@url [0]{\catcode `\\12\catcode `\$12\catcode
  `\&12\catcode `\#12\catcode `\^12\catcode `\_12\catcode `\%12\relax}%
\providecommand \@@startlink[1]{}%
\providecommand \@@endlink[0]{}%
\providecommand \url  [0]{\begingroup\@sanitize@url \@url }%
\providecommand \@url [1]{\endgroup\@href {#1}{\urlprefix }}%
\providecommand \urlprefix  [0]{URL }%
\providecommand \Eprint [0]{\href }%
\providecommand \doibase [0]{http://dx.doi.org/}%
\providecommand \selectlanguage [0]{\@gobble}%
\providecommand \bibinfo  [0]{\@secondoftwo}%
\providecommand \bibfield  [0]{\@secondoftwo}%
\providecommand \translation [1]{[#1]}%
\providecommand \BibitemOpen [0]{}%
\providecommand \bibitemStop [0]{}%
\providecommand \bibitemNoStop [0]{.\EOS\space}%
\providecommand \EOS [0]{\spacefactor3000\relax}%
\providecommand \BibitemShut  [1]{\csname bibitem#1\endcsname}%
\let\auto@bib@innerbib\@empty
\bibitem [{\citenamefont {Wengerowsky}\ \emph {et~al.}(2018)\citenamefont
  {Wengerowsky}, \citenamefont {Joshi}, \citenamefont {Steinlechner},
  \citenamefont {H{\"u}bel},\ and\ \citenamefont
  {Ursin}}]{wengerowsky2018entanglement}%
  \BibitemOpen
  \bibfield  {author} {\bibinfo {author} {\bibfnamefont {S.}~\bibnamefont
  {Wengerowsky}}, \bibinfo {author} {\bibfnamefont {S.~K.}\ \bibnamefont
  {Joshi}}, \bibinfo {author} {\bibfnamefont {F.}~\bibnamefont {Steinlechner}},
  \bibinfo {author} {\bibfnamefont {H.}~\bibnamefont {H{\"u}bel}}, \ and\
  \bibinfo {author} {\bibfnamefont {R.}~\bibnamefont {Ursin}},\ }\href@noop {}
  {\bibfield  {journal} {\bibinfo  {journal} {Nature}\ }\textbf {\bibinfo
  {volume} {564}},\ \bibinfo {pages} {225} (\bibinfo {year}
  {2018})}\BibitemShut {NoStop}%
\bibitem [{\citenamefont {Ursin}\ \emph {et~al.}(2007)\citenamefont {Ursin},
  \citenamefont {Tiefenbacher}, \citenamefont {Schmitt-Manderbach},
  \citenamefont {Weier}, \citenamefont {Scheidl}, \citenamefont {Lindenthal},
  \citenamefont {Blauensteiner}, \citenamefont {Jennewein}, \citenamefont
  {Perdigues}, \citenamefont {Trojek} \emph {et~al.}}]{ursin2007entanglement}%
  \BibitemOpen
  \bibfield  {author} {\bibinfo {author} {\bibfnamefont {R.}~\bibnamefont
  {Ursin}}, \bibinfo {author} {\bibfnamefont {F.}~\bibnamefont {Tiefenbacher}},
  \bibinfo {author} {\bibfnamefont {T.}~\bibnamefont {Schmitt-Manderbach}},
  \bibinfo {author} {\bibfnamefont {H.}~\bibnamefont {Weier}}, \bibinfo
  {author} {\bibfnamefont {T.}~\bibnamefont {Scheidl}}, \bibinfo {author}
  {\bibfnamefont {M.}~\bibnamefont {Lindenthal}}, \bibinfo {author}
  {\bibfnamefont {B.}~\bibnamefont {Blauensteiner}}, \bibinfo {author}
  {\bibfnamefont {T.}~\bibnamefont {Jennewein}}, \bibinfo {author}
  {\bibfnamefont {J.}~\bibnamefont {Perdigues}}, \bibinfo {author}
  {\bibfnamefont {P.}~\bibnamefont {Trojek}},  \emph {et~al.},\ }\href@noop {}
  {\bibfield  {journal} {\bibinfo  {journal} {Nature physics}\ }\textbf
  {\bibinfo {volume} {3}},\ \bibinfo {pages} {481} (\bibinfo {year}
  {2007})}\BibitemShut {NoStop}%
\bibitem [{\citenamefont {Honjo}\ \emph {et~al.}(2008)\citenamefont {Honjo},
  \citenamefont {Nam}, \citenamefont {Takesue}, \citenamefont {Zhang},
  \citenamefont {Kamada}, \citenamefont {Nishida}, \citenamefont {Tadanaga},
  \citenamefont {Asobe}, \citenamefont {Baek}, \citenamefont {Hadfield} \emph
  {et~al.}}]{honjo2008long}%
  \BibitemOpen
  \bibfield  {author} {\bibinfo {author} {\bibfnamefont {T.}~\bibnamefont
  {Honjo}}, \bibinfo {author} {\bibfnamefont {S.~W.}\ \bibnamefont {Nam}},
  \bibinfo {author} {\bibfnamefont {H.}~\bibnamefont {Takesue}}, \bibinfo
  {author} {\bibfnamefont {Q.}~\bibnamefont {Zhang}}, \bibinfo {author}
  {\bibfnamefont {H.}~\bibnamefont {Kamada}}, \bibinfo {author} {\bibfnamefont
  {Y.}~\bibnamefont {Nishida}}, \bibinfo {author} {\bibfnamefont
  {O.}~\bibnamefont {Tadanaga}}, \bibinfo {author} {\bibfnamefont
  {M.}~\bibnamefont {Asobe}}, \bibinfo {author} {\bibfnamefont
  {B.}~\bibnamefont {Baek}}, \bibinfo {author} {\bibfnamefont {R.}~\bibnamefont
  {Hadfield}},  \emph {et~al.},\ }\href@noop {} {\bibfield  {journal} {\bibinfo
   {journal} {Optics Express}\ }\textbf {\bibinfo {volume} {16}},\ \bibinfo
  {pages} {19118} (\bibinfo {year} {2008})}\BibitemShut {NoStop}%
\bibitem [{\citenamefont {Yin}\ \emph {et~al.}(2020)\citenamefont {Yin},
  \citenamefont {Li}, \citenamefont {Liao}, \citenamefont {Yang}, \citenamefont
  {Cao}, \citenamefont {Zhang}, \citenamefont {Ren}, \citenamefont {Cai},
  \citenamefont {Liu}, \citenamefont {Li} \emph
  {et~al.}}]{yin2020entanglement}%
  \BibitemOpen
  \bibfield  {author} {\bibinfo {author} {\bibfnamefont {J.}~\bibnamefont
  {Yin}}, \bibinfo {author} {\bibfnamefont {Y.-H.}\ \bibnamefont {Li}},
  \bibinfo {author} {\bibfnamefont {S.-K.}\ \bibnamefont {Liao}}, \bibinfo
  {author} {\bibfnamefont {M.}~\bibnamefont {Yang}}, \bibinfo {author}
  {\bibfnamefont {Y.}~\bibnamefont {Cao}}, \bibinfo {author} {\bibfnamefont
  {L.}~\bibnamefont {Zhang}}, \bibinfo {author} {\bibfnamefont {J.-G.}\
  \bibnamefont {Ren}}, \bibinfo {author} {\bibfnamefont {W.-Q.}\ \bibnamefont
  {Cai}}, \bibinfo {author} {\bibfnamefont {W.-Y.}\ \bibnamefont {Liu}},
  \bibinfo {author} {\bibfnamefont {S.-L.}\ \bibnamefont {Li}},  \emph
  {et~al.},\ }\href@noop {} {\bibfield  {journal} {\bibinfo  {journal}
  {Nature}\ }\textbf {\bibinfo {volume} {582}},\ \bibinfo {pages} {501}
  (\bibinfo {year} {2020})}\BibitemShut {NoStop}%
\bibitem [{\citenamefont {Nunn}\ \emph {et~al.}(2013)\citenamefont {Nunn},
  \citenamefont {Wright}, \citenamefont {S{\"o}ller}, \citenamefont {Zhang},
  \citenamefont {Walmsley},\ and\ \citenamefont {Smith}}]{nunn2013large}%
  \BibitemOpen
  \bibfield  {author} {\bibinfo {author} {\bibfnamefont {J.}~\bibnamefont
  {Nunn}}, \bibinfo {author} {\bibfnamefont {L.}~\bibnamefont {Wright}},
  \bibinfo {author} {\bibfnamefont {C.}~\bibnamefont {S{\"o}ller}}, \bibinfo
  {author} {\bibfnamefont {L.}~\bibnamefont {Zhang}}, \bibinfo {author}
  {\bibfnamefont {I.}~\bibnamefont {Walmsley}}, \ and\ \bibinfo {author}
  {\bibfnamefont {B.}~\bibnamefont {Smith}},\ }\href@noop {} {\bibfield
  {journal} {\bibinfo  {journal} {Optics express}\ }\textbf {\bibinfo {volume}
  {21}},\ \bibinfo {pages} {15959} (\bibinfo {year} {2013})}\BibitemShut
  {NoStop}%
\bibitem [{\citenamefont {Giovannetti}\ \emph {et~al.}(2004)\citenamefont
  {Giovannetti}, \citenamefont {Lloyd},\ and\ \citenamefont
  {Maccone}}]{SQLlloyd}%
  \BibitemOpen
  \bibfield  {author} {\bibinfo {author} {\bibfnamefont {V.}~\bibnamefont
  {Giovannetti}}, \bibinfo {author} {\bibfnamefont {S.}~\bibnamefont {Lloyd}},
  \ and\ \bibinfo {author} {\bibfnamefont {L.}~\bibnamefont {Maccone}},\
  }\href@noop {} {\bibfield  {journal} {\bibinfo  {journal} {Science}\ }\textbf
  {\bibinfo {volume} {306}},\ \bibinfo {pages} {1330} (\bibinfo {year}
  {2004})}\BibitemShut {NoStop}%
\bibitem [{\citenamefont {Chen}\ \emph {et~al.}(2019)\citenamefont {Chen},
  \citenamefont {Fink}, \citenamefont {Steinlechner}, \citenamefont {Torres},\
  and\ \citenamefont {Ursin}}]{chen2019hong}%
  \BibitemOpen
  \bibfield  {author} {\bibinfo {author} {\bibfnamefont {Y.}~\bibnamefont
  {Chen}}, \bibinfo {author} {\bibfnamefont {M.}~\bibnamefont {Fink}}, \bibinfo
  {author} {\bibfnamefont {F.}~\bibnamefont {Steinlechner}}, \bibinfo {author}
  {\bibfnamefont {J.~P.}\ \bibnamefont {Torres}}, \ and\ \bibinfo {author}
  {\bibfnamefont {R.}~\bibnamefont {Ursin}},\ }\href@noop {} {\bibfield
  {journal} {\bibinfo  {journal} {npj Quantum Information}\ }\textbf {\bibinfo
  {volume} {5}},\ \bibinfo {pages} {1} (\bibinfo {year} {2019})}\BibitemShut
  {NoStop}%
\bibitem [{\citenamefont {Liu}\ \emph {et~al.}(2021)\citenamefont {Liu},
  \citenamefont {Zhang}, \citenamefont {Li}, \citenamefont {Zhang},
  \citenamefont {Yin}, \citenamefont {Fei}, \citenamefont {Li}, \citenamefont
  {Liu}, \citenamefont {Xu}, \citenamefont {Chen} \emph {et~al.}}]{DistPhase}%
  \BibitemOpen
  \bibfield  {author} {\bibinfo {author} {\bibfnamefont {L.-Z.}\ \bibnamefont
  {Liu}}, \bibinfo {author} {\bibfnamefont {Y.-Z.}\ \bibnamefont {Zhang}},
  \bibinfo {author} {\bibfnamefont {Z.-D.}\ \bibnamefont {Li}}, \bibinfo
  {author} {\bibfnamefont {R.}~\bibnamefont {Zhang}}, \bibinfo {author}
  {\bibfnamefont {X.-F.}\ \bibnamefont {Yin}}, \bibinfo {author} {\bibfnamefont
  {Y.-Y.}\ \bibnamefont {Fei}}, \bibinfo {author} {\bibfnamefont
  {L.}~\bibnamefont {Li}}, \bibinfo {author} {\bibfnamefont {N.-L.}\
  \bibnamefont {Liu}}, \bibinfo {author} {\bibfnamefont {F.}~\bibnamefont
  {Xu}}, \bibinfo {author} {\bibfnamefont {Y.-A.}\ \bibnamefont {Chen}},  \emph
  {et~al.},\ }\href@noop {} {\bibfield  {journal} {\bibinfo  {journal} {Nature
  Photonics}\ }\textbf {\bibinfo {volume} {15}},\ \bibinfo {pages} {137}
  (\bibinfo {year} {2021})}\BibitemShut {NoStop}%
\bibitem [{\citenamefont {Lee}\ \emph {et~al.}(2019{\natexlab{a}})\citenamefont
  {Lee}, \citenamefont {Yoon},\ and\ \citenamefont
  {Cho}}]{lee2019interferometric}%
  \BibitemOpen
  \bibfield  {author} {\bibinfo {author} {\bibfnamefont {S.~K.}\ \bibnamefont
  {Lee}}, \bibinfo {author} {\bibfnamefont {T.~H.}\ \bibnamefont {Yoon}}, \
  and\ \bibinfo {author} {\bibfnamefont {M.}~\bibnamefont {Cho}},\ }\href@noop
  {} {\bibfield  {journal} {\bibinfo  {journal} {Optics express}\ }\textbf
  {\bibinfo {volume} {27}},\ \bibinfo {pages} {14853} (\bibinfo {year}
  {2019}{\natexlab{a}})}\BibitemShut {NoStop}%
\bibitem [{\citenamefont {Yabushita}\ and\ \citenamefont
  {Kobayashi}(2004)}]{yabushita2004spectroscopy}%
  \BibitemOpen
  \bibfield  {author} {\bibinfo {author} {\bibfnamefont {A.}~\bibnamefont
  {Yabushita}}\ and\ \bibinfo {author} {\bibfnamefont {T.}~\bibnamefont
  {Kobayashi}},\ }\href@noop {} {\bibfield  {journal} {\bibinfo  {journal}
  {Physical Review A}\ }\textbf {\bibinfo {volume} {69}},\ \bibinfo {pages}
  {013806} (\bibinfo {year} {2004})}\BibitemShut {NoStop}%
\bibitem [{\citenamefont {Lee}\ \emph {et~al.}(2019{\natexlab{b}})\citenamefont
  {Lee}, \citenamefont {Shen}, \citenamefont {Cer{\`e}}, \citenamefont
  {Troupe}, \citenamefont {Lamas-Linares},\ and\ \citenamefont
  {Kurtsiefer}}]{clklee2019symmetrical}%
  \BibitemOpen
  \bibfield  {author} {\bibinfo {author} {\bibfnamefont {J.}~\bibnamefont
  {Lee}}, \bibinfo {author} {\bibfnamefont {L.}~\bibnamefont {Shen}}, \bibinfo
  {author} {\bibfnamefont {A.}~\bibnamefont {Cer{\`e}}}, \bibinfo {author}
  {\bibfnamefont {J.}~\bibnamefont {Troupe}}, \bibinfo {author} {\bibfnamefont
  {A.}~\bibnamefont {Lamas-Linares}}, \ and\ \bibinfo {author} {\bibfnamefont
  {C.}~\bibnamefont {Kurtsiefer}},\ }\href@noop {} {\bibfield  {journal}
  {\bibinfo  {journal} {Applied Physics Letters}\ }\textbf {\bibinfo {volume}
  {114}},\ \bibinfo {pages} {101102} (\bibinfo {year}
  {2019}{\natexlab{b}})}\BibitemShut {NoStop}%
\bibitem [{\citenamefont {Giovannetti}\ \emph
  {et~al.}(2001{\natexlab{a}})\citenamefont {Giovannetti}, \citenamefont
  {Lloyd}, \citenamefont {Maccone},\ and\ \citenamefont
  {Wong}}]{giovannetti2001clock}%
  \BibitemOpen
  \bibfield  {author} {\bibinfo {author} {\bibfnamefont {V.}~\bibnamefont
  {Giovannetti}}, \bibinfo {author} {\bibfnamefont {S.}~\bibnamefont {Lloyd}},
  \bibinfo {author} {\bibfnamefont {L.}~\bibnamefont {Maccone}}, \ and\
  \bibinfo {author} {\bibfnamefont {F.}~\bibnamefont {Wong}},\ }\href@noop {}
  {\bibfield  {journal} {\bibinfo  {journal} {Physical review letters}\
  }\textbf {\bibinfo {volume} {87}},\ \bibinfo {pages} {117902} (\bibinfo
  {year} {2001}{\natexlab{a}})}\BibitemShut {NoStop}%
\bibitem [{\citenamefont {Ho}\ \emph {et~al.}(2009)\citenamefont {Ho},
  \citenamefont {Lamas-Linares},\ and\ \citenamefont
  {Kurtsiefer}}]{ho2009clock}%
  \BibitemOpen
  \bibfield  {author} {\bibinfo {author} {\bibfnamefont {C.}~\bibnamefont
  {Ho}}, \bibinfo {author} {\bibfnamefont {A.}~\bibnamefont {Lamas-Linares}}, \
  and\ \bibinfo {author} {\bibfnamefont {C.}~\bibnamefont {Kurtsiefer}},\
  }\href@noop {} {\bibfield  {journal} {\bibinfo  {journal} {New Journal of
  Physics}\ }\textbf {\bibinfo {volume} {11}},\ \bibinfo {pages} {045011}
  (\bibinfo {year} {2009})}\BibitemShut {NoStop}%
\bibitem [{\citenamefont {Giovannetti}\ \emph
  {et~al.}(2001{\natexlab{b}})\citenamefont {Giovannetti}, \citenamefont
  {Lloyd},\ and\ \citenamefont {Maccone}}]{giovannetti2001quantum}%
  \BibitemOpen
  \bibfield  {author} {\bibinfo {author} {\bibfnamefont {V.}~\bibnamefont
  {Giovannetti}}, \bibinfo {author} {\bibfnamefont {S.}~\bibnamefont {Lloyd}},
  \ and\ \bibinfo {author} {\bibfnamefont {L.}~\bibnamefont {Maccone}},\
  }\href@noop {} {\bibfield  {journal} {\bibinfo  {journal} {Nature}\ }\textbf
  {\bibinfo {volume} {412}},\ \bibinfo {pages} {417} (\bibinfo {year}
  {2001}{\natexlab{b}})}\BibitemShut {NoStop}%
\bibitem [{\citenamefont {Komar}\ \emph {et~al.}(2014)\citenamefont {Komar},
  \citenamefont {Kessler}, \citenamefont {Bishof}, \citenamefont {Jiang},
  \citenamefont {S{\o}rensen}, \citenamefont {Ye},\ and\ \citenamefont
  {Lukin}}]{NetOfClocks}%
  \BibitemOpen
  \bibfield  {author} {\bibinfo {author} {\bibfnamefont {P.}~\bibnamefont
  {Komar}}, \bibinfo {author} {\bibfnamefont {E.~M.}\ \bibnamefont {Kessler}},
  \bibinfo {author} {\bibfnamefont {M.}~\bibnamefont {Bishof}}, \bibinfo
  {author} {\bibfnamefont {L.}~\bibnamefont {Jiang}}, \bibinfo {author}
  {\bibfnamefont {A.~S.}\ \bibnamefont {S{\o}rensen}}, \bibinfo {author}
  {\bibfnamefont {J.}~\bibnamefont {Ye}}, \ and\ \bibinfo {author}
  {\bibfnamefont {M.~D.}\ \bibnamefont {Lukin}},\ }\href@noop {} {\bibfield
  {journal} {\bibinfo  {journal} {Nature Physics}\ }\textbf {\bibinfo {volume}
  {10}},\ \bibinfo {pages} {582} (\bibinfo {year} {2014})}\BibitemShut
  {NoStop}%
\bibitem [{\citenamefont {Spiropulu}\ \emph {et~al.}(2021)\citenamefont
  {Spiropulu}, \citenamefont {Lauk},\ and\ \citenamefont
  {Spentzouris}}]{spiropulu2021illinois}%
  \BibitemOpen
  \bibfield  {author} {\bibinfo {author} {\bibfnamefont {M.}~\bibnamefont
  {Spiropulu}}, \bibinfo {author} {\bibfnamefont {N.}~\bibnamefont {Lauk}}, \
  and\ \bibinfo {author} {\bibfnamefont {P.}~\bibnamefont {Spentzouris}},\
  }\href@noop {} {\bibfield  {journal} {\bibinfo  {journal} {Bulletin of the
  American Physical Society}\ } (\bibinfo {year} {2021})}\BibitemShut {NoStop}%
\bibitem [{\citenamefont {Alshowkan}\ \emph {et~al.}(2021)\citenamefont
  {Alshowkan}, \citenamefont {Williams}, \citenamefont {Evans}, \citenamefont
  {Rao}, \citenamefont {Simmerman}, \citenamefont {Lu}, \citenamefont
  {Lingaraju}, \citenamefont {Weiner}, \citenamefont {Marvinney}, \citenamefont
  {Pai}, \citenamefont {Lawrie}, \citenamefont {Peters},\ and\ \citenamefont
  {Lukens}}]{alshowkan2021reconfigurable}%
  \BibitemOpen
  \bibfield  {author} {\bibinfo {author} {\bibfnamefont {M.}~\bibnamefont
  {Alshowkan}}, \bibinfo {author} {\bibfnamefont {B.~P.}\ \bibnamefont
  {Williams}}, \bibinfo {author} {\bibfnamefont {P.~G.}\ \bibnamefont {Evans}},
  \bibinfo {author} {\bibfnamefont {N.~S.}\ \bibnamefont {Rao}}, \bibinfo
  {author} {\bibfnamefont {E.~M.}\ \bibnamefont {Simmerman}}, \bibinfo {author}
  {\bibfnamefont {H.-H.}\ \bibnamefont {Lu}}, \bibinfo {author} {\bibfnamefont
  {N.~B.}\ \bibnamefont {Lingaraju}}, \bibinfo {author} {\bibfnamefont {A.~M.}\
  \bibnamefont {Weiner}}, \bibinfo {author} {\bibfnamefont {C.~E.}\
  \bibnamefont {Marvinney}}, \bibinfo {author} {\bibfnamefont {Y.-Y.}\
  \bibnamefont {Pai}}, \bibinfo {author} {\bibfnamefont {B.~J.}\ \bibnamefont
  {Lawrie}}, \bibinfo {author} {\bibfnamefont {N.~A.}\ \bibnamefont {Peters}},
  \ and\ \bibinfo {author} {\bibfnamefont {J.~M.}\ \bibnamefont {Lukens}},\
  }\href@noop {} {\bibfield  {journal} {\bibinfo  {journal} {PRX Quantum}\
  }\textbf {\bibinfo {volume} {2}},\ \bibinfo {pages} {040304} (\bibinfo {year}
  {2021})}\BibitemShut {NoStop}%
\bibitem [{\citenamefont {Lingaraju}\ \emph {et~al.}(2021)\citenamefont
  {Lingaraju}, \citenamefont {Lu}, \citenamefont {Seshadri}, \citenamefont
  {Leaird}, \citenamefont {Weiner},\ and\ \citenamefont
  {Lukens}}]{lingaraju2021adaptive}%
  \BibitemOpen
  \bibfield  {author} {\bibinfo {author} {\bibfnamefont {N.~B.}\ \bibnamefont
  {Lingaraju}}, \bibinfo {author} {\bibfnamefont {H.-H.}\ \bibnamefont {Lu}},
  \bibinfo {author} {\bibfnamefont {S.}~\bibnamefont {Seshadri}}, \bibinfo
  {author} {\bibfnamefont {D.~E.}\ \bibnamefont {Leaird}}, \bibinfo {author}
  {\bibfnamefont {A.~M.}\ \bibnamefont {Weiner}}, \ and\ \bibinfo {author}
  {\bibfnamefont {J.~M.}\ \bibnamefont {Lukens}},\ }\href@noop {} {\bibfield
  {journal} {\bibinfo  {journal} {Optica}\ }\textbf {\bibinfo {volume} {8}},\
  \bibinfo {pages} {329} (\bibinfo {year} {2021})}\BibitemShut {NoStop}%
\bibitem [{\citenamefont {Barz}\ \emph {et~al.}(2010)\citenamefont {Barz},
  \citenamefont {Cronenberg}, \citenamefont {Zeilinger},\ and\ \citenamefont
  {Walther}}]{barz2010heralded}%
  \BibitemOpen
  \bibfield  {author} {\bibinfo {author} {\bibfnamefont {S.}~\bibnamefont
  {Barz}}, \bibinfo {author} {\bibfnamefont {G.}~\bibnamefont {Cronenberg}},
  \bibinfo {author} {\bibfnamefont {A.}~\bibnamefont {Zeilinger}}, \ and\
  \bibinfo {author} {\bibfnamefont {P.}~\bibnamefont {Walther}},\ }\href@noop
  {} {\bibfield  {journal} {\bibinfo  {journal} {Nature photonics}\ }\textbf
  {\bibinfo {volume} {4}},\ \bibinfo {pages} {553} (\bibinfo {year}
  {2010})}\BibitemShut {NoStop}%
\bibitem [{\citenamefont {Scherer}\ \emph {et~al.}(2011)\citenamefont
  {Scherer}, \citenamefont {Sanders},\ and\ \citenamefont
  {Tittel}}]{scherer2011long}%
  \BibitemOpen
  \bibfield  {author} {\bibinfo {author} {\bibfnamefont {A.}~\bibnamefont
  {Scherer}}, \bibinfo {author} {\bibfnamefont {B.~C.}\ \bibnamefont
  {Sanders}}, \ and\ \bibinfo {author} {\bibfnamefont {W.}~\bibnamefont
  {Tittel}},\ }\href@noop {} {\bibfield  {journal} {\bibinfo  {journal} {Optics
  express}\ }\textbf {\bibinfo {volume} {19}},\ \bibinfo {pages} {3004}
  (\bibinfo {year} {2011})}\BibitemShut {NoStop}%
\bibitem [{\citenamefont {Sun}\ \emph {et~al.}(2017)\citenamefont {Sun},
  \citenamefont {Jiang}, \citenamefont {Mao}, \citenamefont {You},
  \citenamefont {Zhang}, \citenamefont {Zhang}, \citenamefont {Jiang},
  \citenamefont {Chen}, \citenamefont {Li}, \citenamefont {Huang} \emph
  {et~al.}}]{sun2017entanglement}%
  \BibitemOpen
  \bibfield  {author} {\bibinfo {author} {\bibfnamefont {Q.-C.}\ \bibnamefont
  {Sun}}, \bibinfo {author} {\bibfnamefont {Y.-F.}\ \bibnamefont {Jiang}},
  \bibinfo {author} {\bibfnamefont {Y.-L.}\ \bibnamefont {Mao}}, \bibinfo
  {author} {\bibfnamefont {L.-X.}\ \bibnamefont {You}}, \bibinfo {author}
  {\bibfnamefont {W.}~\bibnamefont {Zhang}}, \bibinfo {author} {\bibfnamefont
  {W.-J.}\ \bibnamefont {Zhang}}, \bibinfo {author} {\bibfnamefont
  {X.}~\bibnamefont {Jiang}}, \bibinfo {author} {\bibfnamefont {T.-Y.}\
  \bibnamefont {Chen}}, \bibinfo {author} {\bibfnamefont {H.}~\bibnamefont
  {Li}}, \bibinfo {author} {\bibfnamefont {Y.-D.}\ \bibnamefont {Huang}},
  \emph {et~al.},\ }\href@noop {} {\bibfield  {journal} {\bibinfo  {journal}
  {Optica}\ }\textbf {\bibinfo {volume} {4}},\ \bibinfo {pages} {1214}
  (\bibinfo {year} {2017})}\BibitemShut {NoStop}%
\bibitem [{\citenamefont {Quan}\ \emph {et~al.}(2020)\citenamefont {Quan},
  \citenamefont {Dong}, \citenamefont {Xiang}, \citenamefont {Li},
  \citenamefont {Liu},\ and\ \citenamefont {Zhang}}]{nonlocalquan2020high}%
  \BibitemOpen
  \bibfield  {author} {\bibinfo {author} {\bibfnamefont {R.}~\bibnamefont
  {Quan}}, \bibinfo {author} {\bibfnamefont {R.}~\bibnamefont {Dong}}, \bibinfo
  {author} {\bibfnamefont {X.}~\bibnamefont {Xiang}}, \bibinfo {author}
  {\bibfnamefont {B.}~\bibnamefont {Li}}, \bibinfo {author} {\bibfnamefont
  {T.}~\bibnamefont {Liu}}, \ and\ \bibinfo {author} {\bibfnamefont
  {S.}~\bibnamefont {Zhang}},\ }\href@noop {} {\bibfield  {journal} {\bibinfo
  {journal} {Review of Scientific Instruments}\ }\textbf {\bibinfo {volume}
  {91}},\ \bibinfo {pages} {123109} (\bibinfo {year} {2020})}\BibitemShut
  {NoStop}%
\bibitem [{\citenamefont {Franson}(1992)}]{fransonnonlocal}%
  \BibitemOpen
  \bibfield  {author} {\bibinfo {author} {\bibfnamefont {J.}~\bibnamefont
  {Franson}},\ }\href@noop {} {\bibfield  {journal} {\bibinfo  {journal}
  {Physical Review A}\ }\textbf {\bibinfo {volume} {45}},\ \bibinfo {pages}
  {3126} (\bibinfo {year} {1992})}\BibitemShut {NoStop}%
\bibitem [{\citenamefont {Fitch}\ and\ \citenamefont
  {Franson}(2002)}]{fitchfransondispersion}%
  \BibitemOpen
  \bibfield  {author} {\bibinfo {author} {\bibfnamefont {M.}~\bibnamefont
  {Fitch}}\ and\ \bibinfo {author} {\bibfnamefont {J.}~\bibnamefont
  {Franson}},\ }\href@noop {} {\bibfield  {journal} {\bibinfo  {journal}
  {Physical Review A}\ }\textbf {\bibinfo {volume} {65}},\ \bibinfo {pages}
  {053809} (\bibinfo {year} {2002})}\BibitemShut {NoStop}%
\bibitem [{\citenamefont {Brendel}\ \emph {et~al.}(1998)\citenamefont
  {Brendel}, \citenamefont {Zbinden},\ and\ \citenamefont {Gisin}}]{chromdisp}%
  \BibitemOpen
  \bibfield  {author} {\bibinfo {author} {\bibfnamefont {J.}~\bibnamefont
  {Brendel}}, \bibinfo {author} {\bibfnamefont {H.}~\bibnamefont {Zbinden}}, \
  and\ \bibinfo {author} {\bibfnamefont {N.}~\bibnamefont {Gisin}},\
  }\href@noop {} {\bibfield  {journal} {\bibinfo  {journal} {Optics
  communications}\ }\textbf {\bibinfo {volume} {151}},\ \bibinfo {pages} {35}
  (\bibinfo {year} {1998})}\BibitemShut {NoStop}%
\bibitem [{\citenamefont {Sensarn}\ \emph {et~al.}(2009)\citenamefont
  {Sensarn}, \citenamefont {Yin},\ and\ \citenamefont {Harris}}]{nonlocalmod}%
  \BibitemOpen
  \bibfield  {author} {\bibinfo {author} {\bibfnamefont {S.}~\bibnamefont
  {Sensarn}}, \bibinfo {author} {\bibfnamefont {G.}~\bibnamefont {Yin}}, \ and\
  \bibinfo {author} {\bibfnamefont {S.}~\bibnamefont {Harris}},\ }\href@noop {}
  {\bibfield  {journal} {\bibinfo  {journal} {Physical review letters}\
  }\textbf {\bibinfo {volume} {103}},\ \bibinfo {pages} {163601} (\bibinfo
  {year} {2009})}\BibitemShut {NoStop}%
\bibitem [{\citenamefont {Korzh}\ \emph {et~al.}(2020)\citenamefont {Korzh},
  \citenamefont {Zhao}, \citenamefont {Allmaras}, \citenamefont {Frasca},
  \citenamefont {Autry}, \citenamefont {Bersin}, \citenamefont {Beyer},
  \citenamefont {Briggs}, \citenamefont {Bumble}, \citenamefont {Colangelo}
  \emph {et~al.}}]{Sub3pskorzh2020demonstration}%
  \BibitemOpen
  \bibfield  {author} {\bibinfo {author} {\bibfnamefont {B.}~\bibnamefont
  {Korzh}}, \bibinfo {author} {\bibfnamefont {Q.-Y.}\ \bibnamefont {Zhao}},
  \bibinfo {author} {\bibfnamefont {J.~P.}\ \bibnamefont {Allmaras}}, \bibinfo
  {author} {\bibfnamefont {S.}~\bibnamefont {Frasca}}, \bibinfo {author}
  {\bibfnamefont {T.~M.}\ \bibnamefont {Autry}}, \bibinfo {author}
  {\bibfnamefont {E.~A.}\ \bibnamefont {Bersin}}, \bibinfo {author}
  {\bibfnamefont {A.~D.}\ \bibnamefont {Beyer}}, \bibinfo {author}
  {\bibfnamefont {R.~M.}\ \bibnamefont {Briggs}}, \bibinfo {author}
  {\bibfnamefont {B.}~\bibnamefont {Bumble}}, \bibinfo {author} {\bibfnamefont
  {M.}~\bibnamefont {Colangelo}},  \emph {et~al.},\ }\href@noop {} {\bibfield
  {journal} {\bibinfo  {journal} {Nature Photonics}\ }\textbf {\bibinfo
  {volume} {14}},\ \bibinfo {pages} {250} (\bibinfo {year} {2020})}\BibitemShut
  {NoStop}%
\bibitem [{\citenamefont {Lyons}\ \emph {et~al.}(2018)\citenamefont {Lyons},
  \citenamefont {Knee}, \citenamefont {Bolduc}, \citenamefont {Roger},
  \citenamefont {Leach}, \citenamefont {Gauger},\ and\ \citenamefont
  {Faccio}}]{lyons2018attosecond}%
  \BibitemOpen
  \bibfield  {author} {\bibinfo {author} {\bibfnamefont {A.}~\bibnamefont
  {Lyons}}, \bibinfo {author} {\bibfnamefont {G.~C.}\ \bibnamefont {Knee}},
  \bibinfo {author} {\bibfnamefont {E.}~\bibnamefont {Bolduc}}, \bibinfo
  {author} {\bibfnamefont {T.}~\bibnamefont {Roger}}, \bibinfo {author}
  {\bibfnamefont {J.}~\bibnamefont {Leach}}, \bibinfo {author} {\bibfnamefont
  {E.~M.}\ \bibnamefont {Gauger}}, \ and\ \bibinfo {author} {\bibfnamefont
  {D.}~\bibnamefont {Faccio}},\ }\href@noop {} {\bibfield  {journal} {\bibinfo
  {journal} {Science advances}\ }\textbf {\bibinfo {volume} {4}},\ \bibinfo
  {pages} {eaap9416} (\bibinfo {year} {2018})}\BibitemShut {NoStop}%
\bibitem [{\citenamefont {Clemmen}\ \emph {et~al.}(2016)\citenamefont
  {Clemmen}, \citenamefont {Farsi}, \citenamefont {Ramelow},\ and\
  \citenamefont {Gaeta}}]{clemmen2016ramsey}%
  \BibitemOpen
  \bibfield  {author} {\bibinfo {author} {\bibfnamefont {S.}~\bibnamefont
  {Clemmen}}, \bibinfo {author} {\bibfnamefont {A.}~\bibnamefont {Farsi}},
  \bibinfo {author} {\bibfnamefont {S.}~\bibnamefont {Ramelow}}, \ and\
  \bibinfo {author} {\bibfnamefont {A.~L.}\ \bibnamefont {Gaeta}},\ }\href@noop
  {} {\bibfield  {journal} {\bibinfo  {journal} {Physical review letters}\
  }\textbf {\bibinfo {volume} {117}},\ \bibinfo {pages} {223601} (\bibinfo
  {year} {2016})}\BibitemShut {NoStop}%
\bibitem [{\citenamefont {Olislager}\ \emph {et~al.}(2010)\citenamefont
  {Olislager}, \citenamefont {Cussey}, \citenamefont {Nguyen}, \citenamefont
  {Emplit}, \citenamefont {Massar}, \citenamefont {Merolla},\ and\
  \citenamefont {Huy}}]{olislager2010frequency}%
  \BibitemOpen
  \bibfield  {author} {\bibinfo {author} {\bibfnamefont {L.}~\bibnamefont
  {Olislager}}, \bibinfo {author} {\bibfnamefont {J.}~\bibnamefont {Cussey}},
  \bibinfo {author} {\bibfnamefont {A.~T.}\ \bibnamefont {Nguyen}}, \bibinfo
  {author} {\bibfnamefont {P.}~\bibnamefont {Emplit}}, \bibinfo {author}
  {\bibfnamefont {S.}~\bibnamefont {Massar}}, \bibinfo {author} {\bibfnamefont
  {J.-M.}\ \bibnamefont {Merolla}}, \ and\ \bibinfo {author} {\bibfnamefont
  {K.~P.}\ \bibnamefont {Huy}},\ }\href@noop {} {\bibfield  {journal} {\bibinfo
   {journal} {Physical Review A}\ }\textbf {\bibinfo {volume} {82}},\ \bibinfo
  {pages} {013804} (\bibinfo {year} {2010})}\BibitemShut {NoStop}%
\bibitem [{\citenamefont {Kues}\ \emph {et~al.}(2017)\citenamefont {Kues},
  \citenamefont {Reimer}, \citenamefont {Roztocki}, \citenamefont {Cort{\'e}s},
  \citenamefont {Sciara}, \citenamefont {Wetzel}, \citenamefont {Zhang},
  \citenamefont {Cino}, \citenamefont {Chu}, \citenamefont {Little} \emph
  {et~al.}}]{kues2017chip}%
  \BibitemOpen
  \bibfield  {author} {\bibinfo {author} {\bibfnamefont {M.}~\bibnamefont
  {Kues}}, \bibinfo {author} {\bibfnamefont {C.}~\bibnamefont {Reimer}},
  \bibinfo {author} {\bibfnamefont {P.}~\bibnamefont {Roztocki}}, \bibinfo
  {author} {\bibfnamefont {L.~R.}\ \bibnamefont {Cort{\'e}s}}, \bibinfo
  {author} {\bibfnamefont {S.}~\bibnamefont {Sciara}}, \bibinfo {author}
  {\bibfnamefont {B.}~\bibnamefont {Wetzel}}, \bibinfo {author} {\bibfnamefont
  {Y.}~\bibnamefont {Zhang}}, \bibinfo {author} {\bibfnamefont
  {A.}~\bibnamefont {Cino}}, \bibinfo {author} {\bibfnamefont {S.~T.}\
  \bibnamefont {Chu}}, \bibinfo {author} {\bibfnamefont {B.~E.}\ \bibnamefont
  {Little}},  \emph {et~al.},\ }\href@noop {} {\bibfield  {journal} {\bibinfo
  {journal} {Nature}\ }\textbf {\bibinfo {volume} {546}},\ \bibinfo {pages}
  {622} (\bibinfo {year} {2017})}\BibitemShut {NoStop}%
\bibitem [{\citenamefont {Imany}\ \emph
  {et~al.}(2018{\natexlab{a}})\citenamefont {Imany}, \citenamefont
  {Jaramillo-Villegas}, \citenamefont {Odele}, \citenamefont {Han},
  \citenamefont {Leaird}, \citenamefont {Lukens}, \citenamefont {Lougovski},
  \citenamefont {Qi},\ and\ \citenamefont {Weiner}}]{imany201850}%
  \BibitemOpen
  \bibfield  {author} {\bibinfo {author} {\bibfnamefont {P.}~\bibnamefont
  {Imany}}, \bibinfo {author} {\bibfnamefont {J.~A.}\ \bibnamefont
  {Jaramillo-Villegas}}, \bibinfo {author} {\bibfnamefont {O.~D.}\ \bibnamefont
  {Odele}}, \bibinfo {author} {\bibfnamefont {K.}~\bibnamefont {Han}}, \bibinfo
  {author} {\bibfnamefont {D.~E.}\ \bibnamefont {Leaird}}, \bibinfo {author}
  {\bibfnamefont {J.~M.}\ \bibnamefont {Lukens}}, \bibinfo {author}
  {\bibfnamefont {P.}~\bibnamefont {Lougovski}}, \bibinfo {author}
  {\bibfnamefont {M.}~\bibnamefont {Qi}}, \ and\ \bibinfo {author}
  {\bibfnamefont {A.~M.}\ \bibnamefont {Weiner}},\ }\href@noop {} {\bibfield
  {journal} {\bibinfo  {journal} {Optics express}\ }\textbf {\bibinfo {volume}
  {26}},\ \bibinfo {pages} {1825} (\bibinfo {year}
  {2018}{\natexlab{a}})}\BibitemShut {NoStop}%
\bibitem [{\citenamefont {Imany}\ \emph
  {et~al.}(2018{\natexlab{b}})\citenamefont {Imany}, \citenamefont {Odele},
  \citenamefont {Jaramillo-Villegas}, \citenamefont {Leaird},\ and\
  \citenamefont {Weiner}}]{imany2018characterization}%
  \BibitemOpen
  \bibfield  {author} {\bibinfo {author} {\bibfnamefont {P.}~\bibnamefont
  {Imany}}, \bibinfo {author} {\bibfnamefont {O.~D.}\ \bibnamefont {Odele}},
  \bibinfo {author} {\bibfnamefont {J.~A.}\ \bibnamefont {Jaramillo-Villegas}},
  \bibinfo {author} {\bibfnamefont {D.~E.}\ \bibnamefont {Leaird}}, \ and\
  \bibinfo {author} {\bibfnamefont {A.~M.}\ \bibnamefont {Weiner}},\
  }\href@noop {} {\bibfield  {journal} {\bibinfo  {journal} {Physical Review
  A}\ }\textbf {\bibinfo {volume} {97}},\ \bibinfo {pages} {013813} (\bibinfo
  {year} {2018}{\natexlab{b}})}\BibitemShut {NoStop}%
\bibitem [{\citenamefont {Imany}\ \emph {et~al.}(2020)\citenamefont {Imany},
  \citenamefont {Lingaraju}, \citenamefont {Alshaykh}, \citenamefont {Leaird},\
  and\ \citenamefont {Weiner}}]{imany2020probing}%
  \BibitemOpen
  \bibfield  {author} {\bibinfo {author} {\bibfnamefont {P.}~\bibnamefont
  {Imany}}, \bibinfo {author} {\bibfnamefont {N.~B.}\ \bibnamefont
  {Lingaraju}}, \bibinfo {author} {\bibfnamefont {M.~S.}\ \bibnamefont
  {Alshaykh}}, \bibinfo {author} {\bibfnamefont {D.~E.}\ \bibnamefont
  {Leaird}}, \ and\ \bibinfo {author} {\bibfnamefont {A.~M.}\ \bibnamefont
  {Weiner}},\ }\href@noop {} {\bibfield  {journal} {\bibinfo  {journal}
  {Science advances}\ }\textbf {\bibinfo {volume} {6}},\ \bibinfo {pages}
  {eaba8066} (\bibinfo {year} {2020})}\BibitemShut {NoStop}%
\bibitem [{\citenamefont {Valencia}\ \emph {et~al.}(2004)\citenamefont
  {Valencia}, \citenamefont {Scarcelli},\ and\ \citenamefont
  {Shih}}]{clkvalencia2004distant}%
  \BibitemOpen
  \bibfield  {author} {\bibinfo {author} {\bibfnamefont {A.}~\bibnamefont
  {Valencia}}, \bibinfo {author} {\bibfnamefont {G.}~\bibnamefont {Scarcelli}},
  \ and\ \bibinfo {author} {\bibfnamefont {Y.}~\bibnamefont {Shih}},\
  }\href@noop {} {\bibfield  {journal} {\bibinfo  {journal} {Applied Physics
  Letters}\ }\textbf {\bibinfo {volume} {85}},\ \bibinfo {pages} {2655}
  (\bibinfo {year} {2004})}\BibitemShut {NoStop}%
\bibitem [{\citenamefont {Seshadri}\ \emph {et~al.}(2020)\citenamefont
  {Seshadri}, \citenamefont {Imany}, \citenamefont {Lingaraju}, \citenamefont
  {Leaird},\ and\ \citenamefont {Weiner}}]{seshadri2020precision}%
  \BibitemOpen
  \bibfield  {author} {\bibinfo {author} {\bibfnamefont {S.}~\bibnamefont
  {Seshadri}}, \bibinfo {author} {\bibfnamefont {P.}~\bibnamefont {Imany}},
  \bibinfo {author} {\bibfnamefont {N.}~\bibnamefont {Lingaraju}}, \bibinfo
  {author} {\bibfnamefont {D.~E.}\ \bibnamefont {Leaird}}, \ and\ \bibinfo
  {author} {\bibfnamefont {A.~M.}\ \bibnamefont {Weiner}},\ }in\ \href@noop {}
  {\emph {\bibinfo {booktitle} {CLEO: QELS\_Fundamental Science}}}\ (\bibinfo
  {organization} {Optical Society of America},\ \bibinfo {year} {2020})\ pp.\
  \bibinfo {pages} {FM1C--5}\BibitemShut {NoStop}%
\bibitem [{\citenamefont {Seshadri}\ \emph {et~al.}(2021)\citenamefont
  {Seshadri}, \citenamefont {Lu}, \citenamefont {Lingaraju}, \citenamefont
  {Imany}, \citenamefont {Leaird},\ and\ \citenamefont
  {Weiner}}]{seshadri2021nonlocal}%
  \BibitemOpen
  \bibfield  {author} {\bibinfo {author} {\bibfnamefont {S.}~\bibnamefont
  {Seshadri}}, \bibinfo {author} {\bibfnamefont {H.-H.}\ \bibnamefont {Lu}},
  \bibinfo {author} {\bibfnamefont {N.~B.}\ \bibnamefont {Lingaraju}}, \bibinfo
  {author} {\bibfnamefont {P.}~\bibnamefont {Imany}}, \bibinfo {author}
  {\bibfnamefont {D.~E.}\ \bibnamefont {Leaird}}, \ and\ \bibinfo {author}
  {\bibfnamefont {A.~M.}\ \bibnamefont {Weiner}},\ }in\ \href@noop {} {\emph
  {\bibinfo {booktitle} {2021 Conference on Lasers and Electro-Optics
  (CLEO)}}}\ (\bibinfo {organization} {IEEE},\ \bibinfo {year} {2021})\ pp.\
  \bibinfo {pages} {1--2}\BibitemShut {NoStop}%
\bibitem [{\citenamefont {Zhu}\ and\ \citenamefont {Wu}(2018)}]{zhu2018dual}%
  \BibitemOpen
  \bibfield  {author} {\bibinfo {author} {\bibfnamefont {Z.}~\bibnamefont
  {Zhu}}\ and\ \bibinfo {author} {\bibfnamefont {G.}~\bibnamefont {Wu}},\
  }\href@noop {} {\bibfield  {journal} {\bibinfo  {journal} {Engineering}\
  }\textbf {\bibinfo {volume} {4}},\ \bibinfo {pages} {772} (\bibinfo {year}
  {2018})}\BibitemShut {NoStop}%
\bibitem [{\citenamefont {Trocha}\ \emph {et~al.}(2018)\citenamefont {Trocha},
  \citenamefont {Karpov}, \citenamefont {Ganin}, \citenamefont {Pfeiffer},
  \citenamefont {Kordts}, \citenamefont {Wolf}, \citenamefont {Krockenberger},
  \citenamefont {Marin-Palomo}, \citenamefont {Weimann}, \citenamefont {Randel}
  \emph {et~al.}}]{trocha2018ultrafast}%
  \BibitemOpen
  \bibfield  {author} {\bibinfo {author} {\bibfnamefont {P.}~\bibnamefont
  {Trocha}}, \bibinfo {author} {\bibfnamefont {M.}~\bibnamefont {Karpov}},
  \bibinfo {author} {\bibfnamefont {D.}~\bibnamefont {Ganin}}, \bibinfo
  {author} {\bibfnamefont {M.~H.}\ \bibnamefont {Pfeiffer}}, \bibinfo {author}
  {\bibfnamefont {A.}~\bibnamefont {Kordts}}, \bibinfo {author} {\bibfnamefont
  {S.}~\bibnamefont {Wolf}}, \bibinfo {author} {\bibfnamefont {J.}~\bibnamefont
  {Krockenberger}}, \bibinfo {author} {\bibfnamefont {P.}~\bibnamefont
  {Marin-Palomo}}, \bibinfo {author} {\bibfnamefont {C.}~\bibnamefont
  {Weimann}}, \bibinfo {author} {\bibfnamefont {S.}~\bibnamefont {Randel}},
  \emph {et~al.},\ }\href@noop {} {\bibfield  {journal} {\bibinfo  {journal}
  {Science}\ }\textbf {\bibinfo {volume} {359}},\ \bibinfo {pages} {887}
  (\bibinfo {year} {2018})}\BibitemShut {NoStop}%
\bibitem [{\citenamefont {Dai}\ \emph {et~al.}(2020)\citenamefont {Dai},
  \citenamefont {Shen}, \citenamefont {Wang}, \citenamefont {Li}, \citenamefont
  {Liu}, \citenamefont {Cai}, \citenamefont {Liao}, \citenamefont {Ren},
  \citenamefont {Yin}, \citenamefont {Chen} \emph {et~al.}}]{dai2020towards}%
  \BibitemOpen
  \bibfield  {author} {\bibinfo {author} {\bibfnamefont {H.}~\bibnamefont
  {Dai}}, \bibinfo {author} {\bibfnamefont {Q.}~\bibnamefont {Shen}}, \bibinfo
  {author} {\bibfnamefont {C.-Z.}\ \bibnamefont {Wang}}, \bibinfo {author}
  {\bibfnamefont {S.-L.}\ \bibnamefont {Li}}, \bibinfo {author} {\bibfnamefont
  {W.-Y.}\ \bibnamefont {Liu}}, \bibinfo {author} {\bibfnamefont {W.-Q.}\
  \bibnamefont {Cai}}, \bibinfo {author} {\bibfnamefont {S.-K.}\ \bibnamefont
  {Liao}}, \bibinfo {author} {\bibfnamefont {J.-G.}\ \bibnamefont {Ren}},
  \bibinfo {author} {\bibfnamefont {J.}~\bibnamefont {Yin}}, \bibinfo {author}
  {\bibfnamefont {Y.-A.}\ \bibnamefont {Chen}},  \emph {et~al.},\ }\href@noop
  {} {\bibfield  {journal} {\bibinfo  {journal} {Nature Physics}\ }\textbf
  {\bibinfo {volume} {16}},\ \bibinfo {pages} {848} (\bibinfo {year}
  {2020})}\BibitemShut {NoStop}%
\bibitem [{\citenamefont {Weiner}(2011)}]{weiner2011ultrafast}%
  \BibitemOpen
  \bibfield  {author} {\bibinfo {author} {\bibfnamefont {A.}~\bibnamefont
  {Weiner}},\ }\href@noop {} {\emph {\bibinfo {title} {Ultrafast optics}}},\
  Vol.~\bibinfo {volume} {72}\ (\bibinfo  {publisher} {John Wiley \& Sons},\
  \bibinfo {year} {2011})\BibitemShut {NoStop}%
\bibitem [{\citenamefont {Pe'Er}\ \emph {et~al.}(2005)\citenamefont {Pe'Er},
  \citenamefont {Dayan}, \citenamefont {Friesem},\ and\ \citenamefont
  {Silberberg}}]{pe2005temporal}%
  \BibitemOpen
  \bibfield  {author} {\bibinfo {author} {\bibfnamefont {A.}~\bibnamefont
  {Pe'Er}}, \bibinfo {author} {\bibfnamefont {B.}~\bibnamefont {Dayan}},
  \bibinfo {author} {\bibfnamefont {A.~A.}\ \bibnamefont {Friesem}}, \ and\
  \bibinfo {author} {\bibfnamefont {Y.}~\bibnamefont {Silberberg}},\
  }\href@noop {} {\bibfield  {journal} {\bibinfo  {journal} {Physical review
  letters}\ }\textbf {\bibinfo {volume} {94}},\ \bibinfo {pages} {073601}
  (\bibinfo {year} {2005})}\BibitemShut {NoStop}%
\bibitem [{\citenamefont {Lukens}\ \emph {et~al.}(2013)\citenamefont {Lukens},
  \citenamefont {Dezfooliyan}, \citenamefont {Langrock}, \citenamefont {Fejer},
  \citenamefont {Leaird},\ and\ \citenamefont {Weiner}}]{lukens2013biphoton}%
  \BibitemOpen
  \bibfield  {author} {\bibinfo {author} {\bibfnamefont {J.~M.}\ \bibnamefont
  {Lukens}}, \bibinfo {author} {\bibfnamefont {A.}~\bibnamefont {Dezfooliyan}},
  \bibinfo {author} {\bibfnamefont {C.}~\bibnamefont {Langrock}}, \bibinfo
  {author} {\bibfnamefont {M.~M.}\ \bibnamefont {Fejer}}, \bibinfo {author}
  {\bibfnamefont {D.~E.}\ \bibnamefont {Leaird}}, \ and\ \bibinfo {author}
  {\bibfnamefont {A.~M.}\ \bibnamefont {Weiner}},\ }\href@noop {} {\bibfield
  {journal} {\bibinfo  {journal} {Optics letters}\ }\textbf {\bibinfo {volume}
  {38}},\ \bibinfo {pages} {4652} (\bibinfo {year} {2013})}\BibitemShut
  {NoStop}%
\end{thebibliography}%

\end{document}